\documentclass[11pt]{article}

\usepackage{graphicx, amssymb, amsmath, fullpage, dsfont, bm, epsfig, multirow,
amsthm, url,color}
\usepackage[citation-order]{amsrefs}

\date{}

\newcommand{\captionfonts}{\small}

\makeatletter  
\long\def\@makecaption#1#2{%
  \vskip\abovecaptionskip
  \sbox\@tempboxa{{\captionfonts #1: #2}}%
  \ifdim \wd\@tempboxa >\hsize
    {\captionfonts #1: #2\par}
  \else
    \hbox to\hsize{\hfil\box\@tempboxa\hfil}%
  \fi
  \vskip\belowcaptionskip}
\makeatother   

\begin{document}

\title{\textbf{Quantum Algorithms for Fermionic Quantum Field Theories}}
\author{{\Large Stephen P.\ Jordan,{\normalsize $^{\dag}$} 
                Keith S.\ M.\ Lee,{\normalsize $^{\ddag\,\sharp}$} 
                and John Preskill {\normalsize $^{\S}$} 
\thanks{\texttt{stephen.jordan@nist.gov}, \texttt{ksml@theory.caltech.edu},
\texttt{preskill@theory.caltech.edu}} } \\[20pt]
\textit{ $^{\dag}$ National Institute of Standards and Technology,
               Gaithersburg, MD, USA } \\[8pt] 
\textit{$^{\ddag}$ Perimeter Institute for Theoretical Physics,
 Waterloo, ON, Canada}\\[8pt]
\textit{$^{\sharp}$
Institute for Quantum Computing and Department of Physics \& Astronomy,}\\[3pt]
\textit{
University of Waterloo, Waterloo, ON, Canada }\\[8pt] 
\textit{$^{\S}$ 
Institute for Quantum Information and Matter,}\\[3pt]
\textit{
California Institute of Technology, Pasadena, CA, USA}}

\bibliographystyle{unsrt}
\maketitle
\newcommand{\ud}{\mathrm{d}}
\newcommand{\braket}[2]{\langle #1|#2\rangle}
\newcommand{\bra}[1]{\langle #1|}
\newcommand{\ket}[1]{|#1\rangle}
\newcommand{\Bra}[1]{\left<#1\right|}
\newcommand{\Ket}[1]{\left|#1\right>}
\newcommand{\Braket}[2]{\left< #1 \right| #2 \right>}
\renewcommand{\th}{^\mathrm{th}}
\newcommand{\tr}{\mathrm{Tr}}
\newcommand{\id}{\mathds{1}}
\newcommand{\eq}[1]{(\ref{#1})}
\newcommand{\sect}[1]{\S\ref{#1}}
\newcommand{\bp}{\mathbf{p}}
\newcommand{\bq}{\mathbf{q}}
\newcommand{\bx}{\mathbf{x}}
\newcommand{\by}{\mathbf{y}}
\newtheorem{theorem}{Theorem}
\newtheorem{proposition}{Proposition}

\begin{abstract}
Extending previous work on scalar field theories, we develop a quantum
algorithm to compute relativistic scattering amplitudes in fermionic
field theories, exemplified by the massive Gross-Neveu model, a theory
in two spacetime dimensions with quartic interactions. The algorithm
introduces new techniques to meet the additional challenges posed by
the characteristics of fermionic fields, and its run time is
polynomial in the desired precision and the energy. Thus, it
constitutes further progress towards an efficient quantum algorithm
for simulating the Standard Model of particle physics.
\end{abstract}

\newpage

\section{Introduction}

Whether a universal quantum computer is sufficiently powerful to be able
to perform quantum field-theoretical computations efficiently has been a
long-standing and important open question.
Efficient quantum algorithms for simulating quantum many-body systems have been
developed theoretically \cite{Lloyd_science,Abrams_Lloyd,Zalka} 
and implemented experimentally \cite{Lanyon:2011,Mueller:2011,Barreiro:2011}, 
but quantum field theory
presents additional technical challenges, such as the formally
infinite number of degrees of freedom per unit volume. In earlier work
\cite{phi4, longversion}, we presented and analyzed a quantum
algorithm for simulating a bosonic quantum field theory called
$\phi^4$ theory. That algorithm runs in a time that is polynomial in
the number of particles, their energy, and the desired precision, and
applies at both weak and strong coupling. Hence, it offers exponential
speedup over existing classical methods at high precision or strong
coupling. In this paper, we extend our work to fermionic quantum field
theories, exemplified by the massive Gross-Neveu model, a theory in
two spacetime dimensions with quartic interactions. Although our
analysis is specific to this theory, our algorithm can be adapted to
other massive fermionic quantum field theories with only minor
modification while retaining polynomial complexity.

Our quantum algorithm generates scattering events: it takes (as the
input) the momenta of the incoming particles and, sampling from the
probability distribution of possible outcomes, returns (as the
output) the momenta of the outgoing particles produced by the
physical scattering process. Physical quantities of interest, such as
scattering cross sections, can thus be approximated by repeated
runs of the simulation, together with statistical data analysis
similar to that used for particle-accelerator experiments.

The features of fermionic field theories not present in bosonic theories pose
new technical problems, the solutions to which require different techniques.
Perhaps the most obvious difference is the anticommutation, rather than
commutation, of fermionic fields.
This forces a change in the representation of the state by qubits: we
use an encoding method for fermionic mode occupation numbers introduced by
Bravyi and Kitaev \cite{Bravyi_Kitaev}. In \cite{longversion}, it was shown 
that simulation of Hamiltonian time evolution via Suzuki-Trotter formulae
has efficiency advantages when applied to spatially local Hamiltonians.
Fermionic anticommutation makes it more difficult to gain efficiency
by exploiting spatial locality. Nevertheless, we obtain a construction
that gives quasi-linear asymptotic scaling in time and the number of
lattice sites, as in the bosonic case.

In contrast with bosonic field theories,
discretization of fermionic field theories leads to the well-known
``fermion doubling'' problem, in which spurious fermion species not in the
continuum theory appear in the discretized theory.
One solution used in lattice gauge theory is to add to the action
the so-called Wilson term, a second-derivative operator that vanishes
in the naive continuum limit. The Wilson term can also
be accommodated in our quantum algorithm; in particular, we show how
it can be turned on during the preparation of the ground state.

In general, state preparation is a demanding task.
The algorithm in \cite{phi4,longversion} uses a three-step procedure.
First, the free vacuum is prepared. For the free scalar theory, this is a
multivariate Gaussian wavefunction.
Next, wavepackets are excited within the free theory. In order that
only single-particle states are created, an ancillary qubit is used,
together with a particular Hamiltonian that acts on the enlarged space.
Finally, the interaction is turned on via a generalization of adiabatic 
state preparation that can be applied to superpositions of eigenstates.
This procedure intersperses backwards time evolutions governed by 
time-independent Hamiltonians into the turn-on to undo the different 
dynamical phases, which otherwise would cause undesirable propagation 
and broadening of wavepackets.

The state-preparation method analyzed here differs from that of
\cite{phi4,longversion} in two main ways.
Preparation of the free vacuum requires modification because the vacuum
of the free fermionic theory is different from that of the free
bosonic theory. For this purpose, we incorporate a separate adiabatic
turn-on step. Furthermore, sources are used to create particle
excitations after the coupling constant is adiabatically turned on,
rather than before. (This difference is not required by the
fermionic nature of the theory.) This method has the advantage that it
works when bound states are possible, in which case the adiabatic
wavepacket preparation of \cite{phi4, longversion} might fail. Another
consequence is that the procedure no longer requires the interleaving
of backwards time evolutions to undo dynamical phases. 
On the other hand, a disadvantage is that the 
preparation of each particle has a significant probability of producing 
no particle. In the case of two-particle scattering, one can perform 
additional repetitions of the simulation, and recognize and discard
simulations in which fewer than two particles have been created. However,
the procedure is not well suited to processes involving
more than two incoming particles.

We analyze two different measurement procedures to be used as the last step
of the simulation. The first method is to return adiabatically to the free
theory and then measure the number operators of the momentum
modes. For unbound states, this procedure yields complete information
about particle momenta, but is not well-suited to detecting bound
states or resolving spatial information. The second procedure is to
measure charge within local regions of space. These measurements can
detect charged bound states, although they are blind to neutral
ones. Which of these measurement schemes is preferable depends on the
desired application.

There is a substantial body of work on analog quantum simulation
of quantum systems, including lattice field theories. (See \cite{Wiese}
for a recent review.) In such work, proposals are made for the
engineering of experimental systems so that they mimic systems of
interest, that is, so that the  Hamiltonians of the laboratory systems
approximate Hamiltonians of interest. The proposed quantum simulators
can be thought of as specialized quantum computers. In contrast, we
address digital quantum algorithms, namely, algorithms to be run on a
universal, fault-tolerant, digital quantum computer. Our work thus
probes the fundamental asymptotic computational complexity of quantum
field theories. 

There is also an extensive literature on the study of quantum field 
theories on classical computers via lattice field theory. 
(See Ch.~17 of \cite{Beringer:1900zz} 
for a review of its results and status.)
However, classical lattice algorithms rely on analytic continuation
to imaginary time, $t \to -i\tau$. Thus, they are useful for 
computing static quantities such as mass ratios, but are unsuitable
for calculating dynamical quantities such as scattering cross
sections. In contrast, our quantum algorithm simulates the dynamics of
quantum field theories, a problem that is expected to be 
$\mathsf{BQP}$-complete and thus impossible to solve by 
polynomial-time classical algorithms.
Although our algorithm draws upon some concepts from lattice field 
theory, new techniques are needed, particularly for state preparation and
measurement. 


The work presented in this paper is another step towards the goal of 
obtaining an efficient quantum algorithm for simulating the Standard Model
of particle physics. Such an algorithm would establish that, except for
quantum-gravity effects, the standard quantum circuit model suffices to
capture completely the computational power of our universe.

The rest of this paper is organized as follows. Section 2 introduces
the massive Gross-Neveu model, gives an overview of our quantum algorithm
for computing the theory's scattering amplitudes, and analyzes
the algorithm's complexity. Section 3 describes in detail the
efficient simulation of the Hamiltonian time evolution in the quantum
circuit model. Section 4 presents our procedures for state preparation
and measurement. Finally, Section 5 addresses some field-theoretical
aspects, namely, the effects of a non-zero lattice spacing and the
renormalization of mass, which are crucial elements in our complexity
analysis.


\section{Quantum Algorithm}

In this section we describe the massive Gross-Neveu model
(\sect{sec:MGN}), outline the steps in our algorithm for simulating
particle scattering processes within this model (\sect{sec:alg}), and
give an overview of the algorithm's complexity
(\sect{sec:complexity}). The run time is polynomial in the inverse of
the desired precision and in the momenta of the incoming particles.
The detailed analysis of the steps of the algorithm that contribute
to the overall complexity stated in \sect{sec:complexity} is given in
later sections.

\subsection{The Massive Gross-Neveu Model}
\label{sec:MGN}

The theory we consider is a generalization of the Gross-Neveu model to
include an explicit mass term in the Lagrangian. The (original) 
Gross-Neveu model \cite{Gross:1974jv} is a quantum field
theory in two spacetime dimensions consisting of
$N$ fermion species with quartic interactions. It has a rich
phenomenology. Like quantum chromodynamics (QCD), the theory governing
the strong interactions, it has the remarkable property of asymptotic
freedom, whereby the interaction becomes weaker at higher
energies. The theory has a discrete chiral symmetry, 
$\psi \rightarrow \gamma^5 \psi$, where 
\begin{equation}
\gamma^5 = \left[ \begin{array}{cc} 1 & 0 \\ 0 & -1
  \end{array} \right].
\end{equation}
This symmetry is spontaneously broken by the
non-perturbative vacuum. (The related theory known as the chiral
Gross-Neveu model has a continuous chiral symmetry, $\psi \rightarrow
e^{i\theta\gamma^5}\psi$.) Correspondingly, mass is generated
dynamically, and the theory admits a topological soliton, the
Callan-Coleman-Gross-Zee (CCGZ) kink. Non-topological
solitons also exist \cite{Dashen:1975xh}.

These interesting characteristics have attracted intense study
and led to applications not only in particle physics but also in
condensed-matter physics, including studies of
ferromagnetic superconductors \cite{Machida:1984zz},
conducting polymers, and systems of strongly correlated electrons
\cite{Lin:1998zz}.

The Gross-Neveu model, together with the chiral Gross-Neveu model,
was originally solved in the limit $N \to \infty$ \cite{Gross:1974jv}.
Via inverse scattering methods \cite{Neveu:1977cr}, and later through
a generalized Bethe Ansatz \cite{Andrei:1979sq}, integrability was
demonstrated for general values of $N$, a feature related to the existence of
infinitely many conserved currents \cite{Brezin:1979am}.
The model's $S$-matrix is factorizable \cite{Zamolodchikov:1978xm,
Karowski:1980kq}: the $n$-body $S$-matrix is expressible as the product of
two-body $S$-matrices.

In contrast, the massive Gross-Neveu model, in which there is an explicit
bare mass, is thought not to be integrable for arbitrary values of $N$.
This theory still exhibits asymptotic freedom, but it does not admit
solitons: for any non-zero mass, the CCGZ kink becomes infinitely massive
and disappears \cite{Feinberg:1996kr}.
The asymptotic freedom and non-zero bare mass make a rigorous perturbative
construction of the theory satisfying the Osterwalder-Schrader axioms
possible \cite{Feldman:1985ar,Feldman:1986ax}.

\smallskip

The massive $N$-component Gross-Neveu model is given by the following
Lagrangian in two spacetime dimensions:
\begin{equation} \label{eq:MGN}
{\cal L} = \sum_{j=1}^N \bar{\psi}_j (i\gamma^\mu \partial_\mu - m) \psi_j 
+ \frac{g^2}{2} \bigg( \sum_{j=1}^N \bar{\psi}_j\psi_j \bigg)^2 \,,
\end{equation}
where each field $\psi_j(x)$ has two components,
$\gamma^\mu$ is a two-dimensional representation of the Dirac algebra,
and $\bar{\psi} = \psi^\dag \gamma^0$.\footnote{
The Dirac matrices satisfy 
$\{\gamma^\mu,\,\gamma^\nu\} \equiv \gamma^\mu\gamma^\nu 
+ \gamma^\nu\gamma^\mu = 2 g^{\mu\nu} \id$, and $\psi_j(x)$ is a spinor,
that is, its Lorentz transformation is such that \eq{eq:MGN} is
Lorentz-invariant. We use the metric $g^{\mu,\nu} = \mathrm{diag}(+1,-1)$.} 
We use the Majorana representation, namely,
\begin{equation}
\label{imp}
\gamma^0 = \left[ \begin{array}{cc} 0 & -i \\
i & 0 \end{array} \right] \,,\quad \quad \gamma^1 =
-\left[ \begin{array}{cc} 0 & i \\ i & 0 \end{array} \right]
\,.
\end{equation}
The components of the field operator associated with the particle species
$j \in \{1,2,\ldots,N\}$ will be denoted by $\psi_{j,\alpha}$,
$\alpha \in \{0,1\}$.
In units where $\hbar = c = 1$, any quantity has units of some power of 
mass, referred to as the mass dimension.
We shall use bold-face to represent spatial vectors, such as $\mathbf{p}$ 
and $\mathbf{x}$, to distinguish them from spacetime vectors $x^\mu =
(t,\mathbf{x})$ and $p^\mu = (E,\mathbf{p})$. Note, however, that we are
considering 1+1 dimensions; thus, spatial vectors have only one
component. 

The dimensionless parameter $g$ determines the strength of the interaction. 
When $g=0$, the $\psi_j$ are free fields obeying the Dirac equation,
$(i\gamma^\mu \partial_\mu - m_0) \psi_j(x) = 0$. 
%
Then one can write
\begin{equation}
\psi_j (x) = \int\frac{d\bp}{2\pi}\frac{1}{\sqrt{2 E_{\bp}}} 
\left( a_j(\bp) u(\bp) e^{-ip\cdot x}
 + b_j^\dag(\bp) v(\bp) e^{ip\cdot x} \right)\,, 
   \label{eq:psi}
\end{equation}
where 
\begin{equation}
E_{\bp} = \sqrt{\bp^2 + m_0^2} \,,
\end{equation}
$ a_j(\bp),\, b_j^\dag(\bp)$ are creation and annihilation operators,
and $u,v$ satisfy
\begin{eqnarray}
(m_0 \gamma^0 + \bp \gamma^0 \gamma^1) u(\bp) & = &
  E_{\bp} u(\bp) \,, \label{ident1}\\
(m_0 \gamma^0 - \bp \gamma^0 \gamma^1 ) v(\bp) & = & -
  E_{\bp} v(\bp) \,, \label{ident2}\\
u^\dag(\bp) u(\bp) = v^\dag(\bp) v(\bp) 
& = & 2 E_{\bp} \,, \label{ident3}\\
u(\bp)^\dag v(-\bp) & = & 0 \,, \label{ident4}\\
\bar{u}(\bp) u(\bp) = - \bar{v}(\bp)
v(\bp) & = & 2 m_0 \,, \label{ident5}\\
\bar{u}(\bp) v(\bp) = \bar{v}(\bp)
u(\bp) & = & 0 \,. \label{ident6}
\end{eqnarray}
In the Majorana representation \eq{imp}, one 
has the following concrete solution:
\begin{equation}
\label{concrete}
u(\bp) = \left[ \begin{array}{c} \sqrt{E_{\bp} -
      \bp} \\
i \sqrt{E_{\bp} + \bp} \end{array} \right] \, \,, \quad
v(\bp) = \left[ \begin{array}{c} \sqrt{E_{\bp} -
      \bp} \\
-i \sqrt{ E_{\bp} + \bp} \end{array} \right] \,.
\end{equation}
%


\subsection{Description of Algorithm}
\label{sec:alg}

To represent the field using qubits, we first discretize the quantum
field theory, putting it on a spatial lattice. (Discretization errors are
analyzed in \sect{EFT}.) Having done that, our algorithm consists of six
main steps, which we analyze in subsequent sections.

\begin{enumerate}
\item Prepare the ground state of the Hamiltonian with
  both the interaction term ($g_0^2$) and the
  nearest-neighbor lattice-site interactions turned off. This can be
  done efficiently because the ground state is a tensor product of the
  ground states of the individual lattice sites.
\item Simulate, via Suzuki-Trotter formulae, the adiabatic turn-on of
  the nearest-neighbor lattice-site interactions, thereby obtaining
  the ground state of the non-interacting theory.
\item Adiabatically turn on the interaction term, while
  adjusting the parameter $m_0$ to compensate for the renormalization
  of the physical mass.
\item Excite particle wavepackets, by introducing a
  source term in the Hamiltonian. The source term is chosen to be
  sinusoidally varying in time and space so as to select the desired
  mass and momentum of particle excitations by resonance.
\item Evolve in time, via Suzuki-Trotter formulae,
  according to the full massive Gross-Neveu Hamiltonian. It is during this
  time evolution that scattering may occur.
\item Either use phase estimation to measure local charge observables,
  or adiabatically return to the free theory and then use phase
  estimation to measure number operators of momentum modes. (The
  choice between these forms of measurement depends on the application.)
\end{enumerate}


\subsection{Complexity}
\label{sec:complexity}

In this section we bound the asymptotic scaling of the number of gates
needed to simulate scattering processes as a function of the momentum
$p$ of the incoming particles and the precision $\epsilon$ to which
the final results are desired. The effect of discretization, via a 
lattice of spacing $a$, is captured by (infinitely many) terms in the
effective Hamiltonian that are not present in the continuum massive 
Gross-Neveu theory (\sect{EFT}). Truncation of these terms, 
which make contributions of $O(a)$ to scattering cross sections,
therefore constitutes an error. Thus, 
to ensure any cross section $\sigma'$ in the discretized quantum field
theory matches the continuum value $\sigma$ to within
\begin{equation}
\label{criterion}
(1-\epsilon) \sigma \leq \sigma' \leq (1+\epsilon) \sigma,
\end{equation}
one must choose the scaling $a \sim \epsilon$ in the high-precision limit, 
that is, the limit $\epsilon \to 0$. Similarly, in the large-momentum limit, 
one must choose the scaling $a \sim p^{-1}$ in order to ensure that the 
wavelength of each particle is large compared with the lattice spacing.

It suffices to use an adiabatic
process of duration
\begin{equation}
T = O \left( \frac{L^2}{a^4 m^3 \epsilon} \right)
\end{equation}
(where $L$ is the length of the spatial dimension and $m$ is the physical
mass)
to prepare a state within a distance $\epsilon$ of the free vacuum 
(\sect{freeprep}). 
Using Suzuki-Trotter decompositions of the form described in \sect{trotter},
we can simulate this adiabatic time evolution using a number of quantum gates
scaling as
\begin{eqnarray}
G_{\mathrm{prep}} & = & O\left( \left( \frac{TL}{a^2} \right)^{1+o(1)}
\epsilon^{-o(1)} \right) \\
& = & O \left( \left( \frac{L^3}{a^6 m^3 \epsilon} \right)^{1+o(1)}
\right) \,. \label{gprep}
\end{eqnarray}

The next state-preparation step is to simulate adiabatic turn-on of
the coupling, thereby obtaining the interacting vacuum.
This can be achieved in a time (\sect{turnon})
\begin{equation}
T_{\mathrm{turn-on}} = O \left( \frac{L^2}{a^4 m^3 \epsilon }
\right).
\end{equation}
Applying Suzuki-Trotter formulae, one obtains a gate count of
\begin{equation}
G_{\mathrm{turn-on}} = O \left( \left( \frac{L^3}{a^6 m^3 \epsilon}
\right)^{1+o(1)} \right). \label{gturnon}
\end{equation}

The final state-preparation step is to excite particle wavepackets. We
do this by applying a time-dependent perturbation $\lambda W(t)$ for
time $\tau$. It is necessary to choose $\tau$ large enough and
$\lambda$ small enough to suppress the production of particle
pairs. The choice of small $\lambda$ means that there will be a substantial
probability that no particle is produced. Let $p_1$ denote the probability
that exactly one particle is produced. In a typical simulation one
wishes to produce an initial state of two spatially separated incoming
particles. The probability that both of these are produced is
$p_1^2$. The simulations in which one or both initial particles has failed 
to be created can be detected at the final measurement stage of the
simulation and discarded. This comes at the cost of a factor of
$1/p_1^2$ more repetitions of the simulation. 
The probability $p_1$ is independent of momentum and scales with 
precision as $p_1 \sim \epsilon$ (\sect{exciting}). 
Also, in \sect{exciting} one finds that the total
number of quantum gates needed for the excitation step is
\begin{equation}
G_{\mathrm{excite}} = \left\{ \begin{array}{ll}
\epsilon^{-4-o(1)} \,, & \textrm{as} \,\,\, \epsilon \to 0 \,, \\
p^{3+o(1)} \,, & \textrm{as} \,\,\, p \to \infty \,.
\end{array} \right.
\end{equation}

In both the high-momentum and high-precision limits, the dominant costs in
the algorithm are the two adiabatic state preparation steps, whose
complexity is given in \eq{gprep} and \eq{gturnon}.
In the high-precision limit, to compute physical quantities such as
scattering cross sections to within a factor of $(1+\epsilon)$, one
must choose $a$ to scale as $\epsilon$ (\sect{EFT}). Also, in this limit,
the complexity contains a further factor of $1/\epsilon$ owing to
postselection of simulations in which both wavepacket excitations have been
successful (\sect{exciting}). 
Substituting
$a \sim \epsilon$ into \eq{gprep} and including this extra factor of
$1/\epsilon$ yield a total complexity of
$O(\epsilon^{-8-o(1)})$. 
In the  high-momentum limit, $a$ must
scale as $1/p$ to ensure that the particle wavelength is long compared
to the lattice spacing, and $L$ must scale as $p$ to accommodate the
excitation step (\sect{exciting}). 
In summary, we obtain
\begin{equation}
G_{\mathrm{total}} = \left\{ \begin{array}{ll}
O(\epsilon^{-8-o(1)}) \,, & \textrm{as} \,\,\, \epsilon \to 0 \,, \\
O(p^{9+o(1)}) \,, & \textrm{as} \,\,\,  p \to \infty \,. \end{array} \right.
\end{equation}
Note that these are only upper bounds on the complexity, and it may be
possible to improve them by using more detailed analysis, such as more
specialized adiabatic theorems.


\section{Qubits and Quantum Gates}

We divide the problem of simulating Hamiltonian time evolutions
in the massive Gross-Neveu model into three subproblems. 
The first subproblem is to represent the state of the field with qubits. 
We do this by choosing a complete set of commuting observables and encoding 
their eigenvalues with strings of bits (\sect{rep}). The second subproblem
is to simulate local fermionic gates on the degrees of freedom defined by
the commuting observables. Achieving this in an efficient manner is
non-trivial because of the fermionic statistics. For this purpose, we employ
a technique due to Bravyi and Kitaev \cite{Bravyi_Kitaev}, which
implements fermionic statistics with only logarithmic overhead in the
number of lattice sites (\sect{BK}). The third subproblem is to decompose the
time evolution governed by the massive Gross-Neveu Hamiltonian into a product
of local fermionic gates. We do this using high-order
Suzuki-Trotter formulae \cite{Suzuki90} with optimizations tailored to
the fermionic statistics and the spatially local nature of the Hamiltonian 
(\sect{trotter}). The local unitary transformations act on at most
$2^{2N}$-dimensional Hilbert spaces and can therefore be efficiently
decomposed into elementary gates for any constant number of particle
species, $N$, via the Solovay-Kitaev algorithm \cite{Kitaev97, Dawson_Nielsen}.


\subsection{Representation by Qubits}
\label{rep}

First, we put the massive Gross-Neveu model on a spatial lattice
\begin{equation}
\Omega = a \mathbb{Z}_{\hat{L}}  \,.
\end{equation}
For simplicity, we impose periodic
boundary conditions, so that $\Omega$ can be considered a circle of
circumference $L = a \hat{L}$. 
The Hamiltonian is
\begin{equation}
\label{h}
H = H_0 + H_g + H_W \,,
\end{equation}
where
\begin{eqnarray}
H_0 & = & \sum_{\bx \in \Omega} a \sum_{j=1}^N \bar{\psi}_j(\bx) 
\left[ -i \gamma^1 \frac{\psi_j(\bx + a)
- \psi_j(\bx-a)}{2a} + m_0 \psi_j(\bx) \right] \label{h0} \,, \\
H_g & = & -\frac{g_0^2}{2} \sum_{\bx \in \Omega} a \bigg( \sum_{j=1}^N
\bar{\psi}_j (\bx) \psi_j(\bx) \bigg)^2 \label{hg} \,, \\
H_W & = & \sum_{\bx \in \Omega} a \sum_{j=1}^N \left[ - \frac{r}{2a} 
\bar{\psi}_j(\bx) \left( \psi_j(\bx+a) - 2 \psi_j(\bx) + \psi_j(\bx-a) \right) 
\right] \,. 
\label{hw} 
\end{eqnarray}
Here, $H_g$ is the interaction term, and $H_W$ is the Wilson term, used to
prevent fermion doubling \cite{Wilson74}. Correspondingly, $0 < r
\leq 1$ is called the Wilson parameter. $H$ is spatially local in the sense 
that it consists only of single-site and nearest-neighbor terms on the
lattice.

Let $\Gamma$ denote the momentum-space lattice corresponding to
$\Omega$, namely,
\begin{equation}
\Gamma = \frac{2\pi}{L} \mathbb{Z}_{\hat{L}} \,.
\end{equation}
We can deduce the spectrum $H_0 + H_W$ using 
\begin{eqnarray}
\label{psij}
\psi_j(\bx) & = & \sum_{\bp \in \Gamma} \frac{1}{L}
\frac{1}{\sqrt{2 E_{\bp}}} \left( a_j(\bp) u(\bp) e^{i \bp \cdot \bx}
  + b_j^\dag(\bp) v(\bp) e^{-i \bp \cdot \bx} \right) \,,\\
\label{psidagj}
\bar{\psi}_j(\bx) & = & \sum_{\bp \in \Gamma} \frac{1}{L}
\frac{1}{\sqrt{2 E_{\bp}}} \left( a_j^\dag(\bp) \bar{u}(\bp) e^{-i \bp
  \cdot \bx} + b_j(\bp) \bar{v}(\bp) e^{i \bp \cdot \bx} \right) \,.
\end{eqnarray}
The inverse transformation is
\begin{eqnarray}
a_j(\bp) & = & \frac{1}{\sqrt{2 E_{\bp}}} u^\dag(\bp) \sum_{\bx \in \Omega}
a e^{-i \bp \cdot \bx} \psi_j(\bx) \,, \label{adef}\\
b_j^\dag(\bp) & = & \frac{1}{\sqrt{2 E_{\bp}}} v^\dag(\bp) \sum_{\bx
  \in \Omega} a e^{i \bp \cdot \bx} \psi_j(\bx) \label{bdef}\,.
\end{eqnarray}
Substituting \eq{psij} and \eq{psidagj} into \eq{h0} and \eq{hw} and
neglecting the vacuum energy, we obtain
\begin{equation}
H_0 + H_W = \sum_{j=1}^N \sum_{\bp \in \Gamma} \frac{1}{L} E^{(a)}_{\bp}(m_0)
\left( a_j^\dag(\bp) a_j(\bp) + b_j^\dag(\bp) b_j(\bp) \right) \,,
\end{equation}
where
\begin{equation}
E^{(a)}_{\bp}(m_0) = \sqrt{\left( m_0 + \frac{2r}{a} \sin^2 \left(
  \frac{\bp a}{2} \right) \right)^2 + \frac{1}{a^2} \sin^2 (\bp a)} \,.
\label{eq:Epa}
\end{equation}
From the canonical fermionic anticommutation relations
\begin{eqnarray}
\label{anticanon1}
\{ \psi_{j,\alpha}(\bx), \psi^\dag_{k,\beta}(\by)\} & = & a^{-1}
\delta_{\bx, \by} \delta_{j,k} \delta_{\alpha, \beta} \id \,, \\
\label{anticanon2}
\{ \psi^\dag_{j,\alpha}(\bx), \psi^\dag_{k,\beta}(\by) \} & = &
\{\psi_{j,\alpha}(\bx), \psi_{k,\beta}(\by) \}  =  0 \,,
\end{eqnarray}
it follows that
\begin{eqnarray}
\label{anticanonp1}
\{ a_j(\bp), a_k^\dag(\bq) \} & = & L \delta_{\bp, \bq} \delta_{j,k}
\id \,, \\
\label{anticanonp2}
\{ b_j(\bp), b_k^\dag(\bq) \} & = & L \delta_{\bp, \bq} \delta_{j,k}
\id \,,
\end{eqnarray}
with all other anticommutators involving $a$ and $b$ operators equal to
zero. We thus have the following interpretation: there are $N$
independent fermion species, created (with momentum $\bp$) by
$a_1^\dag(\bp),\ldots,a_N^\dag(\bp)$ and annihilated by
$a_1(\bp),\ldots,a_N(\bp)$. Similarly, for each species $j$, 
$b_j^\dag(\bp)$ and $b_j(\bp)$ are the creation and annihilation
operators for a corresponding antifermion. Thus, $H$ acts
on a Hilbert space of dimension $2^{2N\hat{L}}$.

\medskip

We can specify a basis for the Hilbert space of field states by
choosing a complete set of commuting observables. The basis is then
indexed by the set of eigenvalues of these observables. 
The fermionic anticommutation relations $ \{a,a^\dag\} = \id ,\,
\{a,a\} = 0$ imply that the algebra generated by $a$ and $a^\dag$ has
the irreducible representation
$
a \to \left[ \begin{array}{cc} 0 & 1 \\ 0 & 0 \end{array} \right]
\,,\,
a^\dag \to \left[ \begin{array}{cc} 0 & 0 \\ 1 & 0 \end{array} \right]
$,
which is unique up to the choice of basis. Hence, the eigenvalues of
$a^\dag a$ are $0$ and $1$. The two basis vectors for the space on
which $a$ and $a^\dag$ act are interpreted as the presence or absence
of a fermion.
%

Thus, by \eq{anticanon1} and \eq{anticanon2},
\begin{equation}
S_x = \{a \psi^\dag_{j,\alpha}(\mathbf{x})
\psi_{j,\alpha}(\mathbf{x})|j=1,\ldots,N; \ \alpha=0,1; \ 
\mathbf{x} \in \Omega \}
\end{equation}
is a set of $2N\hat{L}$ commuting observables, each of which has
eigenvalues zero and one. Similarly, by \eq{anticanonp1} and
\eq{anticanonp2}, 
\begin{equation}
S_p = \{ L^{-1} a_j^\dag(\mathbf{p}) a_j(\mathbf{p})| j=1,\ldots,N;
\ \mathbf{p} \in \Gamma \} \cup \{ L^{-1} b_j^\dag(\mathbf{p})
b_j(\mathbf{p})| j=1,\ldots,N; \ \mathbf{p} \in \Gamma \}
\end{equation}
is a set of $2N\hat{L}$ commuting observables, each with eigenvalues
zero and one. In the non-interacting theory, the eigenvalues of the
elements of $S_p$ are interpreted as the fermionic occupation numbers 
of different momentum modes.

The Hamiltonian $H_0+H_W$ is called the free theory. The
eigenstates of the number operators in $S_p$ are eigenstates of
$H_0+H_W$, and thus the particles do not interact. The rest mass
of these non-interacting particles is $E_0^{(a)}(m_0) =
m_0$. It is not known how to solve for the spectrum of
$H_0+H_W+H_g$ analytically, but the eigenvalue spectrum of
$H_0+H_W+H_g$ can still be characterized in terms of particles. The
rest mass $m$ of the particles in $H_0+H_W+H_g$ is equal to the
eigenvalue gap between the ground state (also called the vacuum) and
the first excited state. In the interacting theory, it is no longer
true that $m = m_0$. Rather, $m$ depends in a non-trivial way on $m_0$,
$g_0$, and $a$; the mass is said to be renormalized. 
A quantitative analysis of this effect contributes to our
analysis of adiabatic state preparation and is given in
\sect{massren}.

One can represent the quantum state of the fermionic fields using
$2N\hat{L}$ qubits to store the eigenvalues of the elements of either
$S_x$ or $S_p$. The ground state of
the free theory in the $S_p$ representation is thus 
$\ket{000\ldots}$. However, the ground state of the interacting theory
is non-trivial in both 
representations. We define our qubit basis in terms of the elements of
$S_x$, because the Gross-Neveu Hamiltonian is local in this basis,
which improves the scaling of the Suzuki-Trotter formulae used to
implement time evolution. 
However, we do not simply store the eigenvalues of the
elements of $S_x$ directly as the values of the qubits. 
This representation would be somewhat inefficient to act upon, because 
direct implementation of the fermionic minus signs requires $O(\hat{L})$ 
gates. Instead,
we apply the method of \cite{Bravyi_Kitaev} to reduce this overhead to
$O(\log \hat{L})$, as described next. 


\subsection{Simulating Fermionic Gates}
\label{BK}

The implementation of fermionic gates using qubits can present a technical
challenge \cite{Bravyi_Kitaev}. As an example, consider the unitary 
transformation $U_{j,\alpha}(\mathbf{x}) = \sqrt{a} \big(
  \psi_{j,\alpha}(\mathbf{x}) + \psi^\dag_{j,\alpha}(\mathbf{x})
\big)$. This toggles the eigenvalue of $a
\psi_{j,\alpha}(\mathbf{x}) \psi^\dag_{j,\alpha}(\mathbf{x})$ between
zero and one. Such a toggling can be implemented on qubits  with the
NOT gate. However, to satisfy the fermionic anticommutation relations
\eq{anticanon1} and \eq{anticanon2} the sign of the transition
amplitude between the zero and one state must depend on the occupation
of other modes. A well-known way to satisfy
\eq{anticanon1} and \eq{anticanon2} is to use a Jordan-Wigner
transformation, in which the modes are given an ordering and
$U_{j,\alpha}(\mathbf{x})$ is represented by the operator $\sigma_x
\otimes \sigma_z \otimes \ldots \otimes \sigma_z$, where the
$\sigma_z$ operators apply to all preceding modes\footnote{Note that
  one can apply both the Jordan-Wigner and Bravyi-Kitaev methods for
  implementing fermionic operators on quantum computers
  in any number of spatial dimensions, using an arbitrary numbering of
  modes.
}
\cite{Jordan_Wigner}. Unfortunately, this method clearly has an $O(\hat{L})$
overhead. In \cite{Bravyi_Kitaev}, Bravyi and Kitaev give a method 
with only $O(\log \hat{L})$ overhead, which we briefly review
here. 

Let $n_i$ be the occupation number of the $i\th$ fermionic mode according 
to some chosen numbering of the modes from 1 to $2N\hat{L}$. To implement
the minus signs in $U_{j,\alpha}(\mathbf{x})$, one needs to know
$\sum_i n_i$, where the sum is over all preceding modes. Thus, a
natural encoding of fermionic mode occupation numbers is to store
the quantities $t_i = \sum_{j=1}^i n_j$ instead of the quantities
$n_i$. This encoding has the advantage that calculating the relevant signs has
an $O(1)$ cost. However, it has the disadvantage that, if the
occupation number of the $i\th$ mode changes, then $i-1$ of the $t_i$ values
must be updated. Thus, updates have an $O(\hat{L})$ cost. 
The Bravyi-Kitaev encoding uses the following compromise, 
in which the calculation of the relevant signs and the
update steps can both be performed in time $O(\log \hat{L})$.

The mode index $i \in \{1,\ldots,2N\hat{L}\}$ can be represented by a
bit string of length $l = \lceil \log_2 (2N\hat{L}) \rceil$. One can
define the following partial order on these bit strings. Consider two
bit strings $x = x_l x_{l-1} \ldots x_1$ and $y = y_l y_{l-1} \ldots
y_1$. Then $x \preceq y$ if, for some $r$, $x_j = y_j$ for $j > r$ and
$y_{r-1} = y_{r-2} = \ldots = y_1 = 1$. Now, let $k_j = \sum_{s
  \preceq j} n_s$. Any total
occupation number $t_i$ can be computed from the $k_j$ quantities in $O(\log
\hat{L})$ time and changing the occcupation of any mode $n_j$
requires updating only $O(\log \hat{L})$ of the $k_j$ quantities 
\cite{Bravyi_Kitaev}.

In fact, the Bravyi-Kitaev construction is
relevant only to the excitation of wavepackets 
(\sect{exciting}). In all other parts of our algorithm, we simulate a
Hamiltonian in which every term is a product of an even number of fermionic
field operators, all acting on the same site or on
nearest-neighbor sites in one dimension. In this case, traditional
Jordan-Wigner techniques incur only $O(1)$ overhead.


\subsection{Application of Suzuki-Trotter Formulae to Fermionic systems}
\label{trotter}

In this section, we describe how to construct efficient quantum
circuits that simulate time evolution induced by the Hamiltonian $H$
defined in \eq{h}, \eq{h0}, \eq{hg}, and
\eq{hw}. We present the case in which $H$ is time-independent. By the results
of \cite{Suzuki93}, the same analysis applies to the simulation of the
time-dependent Hamiltonians that we use in adiabatic state
preparation. (See also \cite{Wiebe}.)

Using a $k^{\mathrm{th}}$-order Suzuki-Trotter formula, one can implement
Hamiltonian time evolution of duration $t$ using a number of quantum
gates that scales as $t^{1+\frac{1}{2k}}$ \cite{Suzuki90, Cleve_sim}. 
Generally, applying a Suzuki-Trotter formula directly to a Hamiltonian of 
the form
\begin{equation}
\label{manyterms}
H = \sum_{i=1}^m H_i
\end{equation}
yields an algorithm with $O(m^{1+o(1)})$ timesteps, and
hence $O(m^{2+o(1)})$ gates, if the $H_i$ are not mutually commuting. 
Thus, 
it is often
advantageous to group terms in a Hamiltonian like \eq{manyterms}
into as small a collection as possible of sets of mutually
commuting terms \cite{Raeisi,longversion}. 

Consider the problem of simulating the Hamiltonian $H$ defined in
\eq{h}, \eq{h0}, \eq{hg}, and \eq{hw}. By
\eq{anticanon1} and \eq{anticanon2}, one
sees that 
\begin{equation}
\label{sitecom}
[\bar{\psi}_j(\mathbf{x}) \psi_j(\mathbf{x}), \bar{\psi}_k(\mathbf{y})
\psi_k(\mathbf{y})] = 0 \,,
\end{equation}
regardless of whether $j=k$ or $\mathbf{x}=\mathbf{y}$. Thus, we start by
decomposing $H$ as a sum of two parts, the single-site terms and the
terms that couple nearest neighbors:
\begin{equation}
H = H_{\mathrm{ss}} + H_{\mathrm{nn}} \,,
\end{equation}
where
\begin{equation}
H_{\mathrm{ss}} = \sum_{\mathbf{x} \in \Omega} a \Bigg[ \sum_{j=1}^N
  \left( m_0 \bar{\psi}_j(\mathbf{x}) \psi_j(\mathbf{x}) + \frac{r}{a}
    \bar{\psi}_j(\mathbf{x}) \psi_j(\mathbf{x}) \right) +
  \frac{g_0^2}{2} \bigg( \sum_{j=1}^N \bar{\psi}_j(\mathbf{x})
    \psi_j(\mathbf{x}) \bigg)^2 \Bigg] \,.
\end{equation}
By \eq{sitecom}, $e^{-i H_{\mathrm{ss}} \delta t}$ decomposes into a
product of local unitary transformations.

All terms in $H_{\mathrm{nn}}$ are of the form
\begin{equation}
\label{form}
\psi_{j,\alpha}^\dag (\mathbf{x}) \psi_{j, \beta}(\mathbf{y}) +
\psi^\dag_{j, \beta}(\mathbf{y}) \psi_{j,\alpha}(\mathbf{x}) \,,
\end{equation}
for $\mathbf{x} = \mathbf{y} \pm a$. Terms with $\alpha = \beta$ and
terms with $\alpha \neq \beta$ are both present in $H_{\mathrm{nn}}$.

Given an operator of the form \eq{form}, let us refer to the
subset of $\{1,\ldots,N\} \times \{0,1\} \times \Omega$ on which it
acts as its support. Because they consist of a product of an even number 
of fermionic operators, any two operators of the form \eq{form} commute
provided they have disjoint support. Thus, we next decompose
$H_{\mathrm{nn}}$ as
\begin{equation}
\label{fourcolors}
H_{\mathrm{nn}} = H_1 + H_2 + H_3 + H_4 \,,
\end{equation}
where each of $H_1,\ldots,H_4$ consists of a sum of terms with
non-intersecting support.

\begin{figure}
\begin{center}
\includegraphics[width=0.5\textwidth]{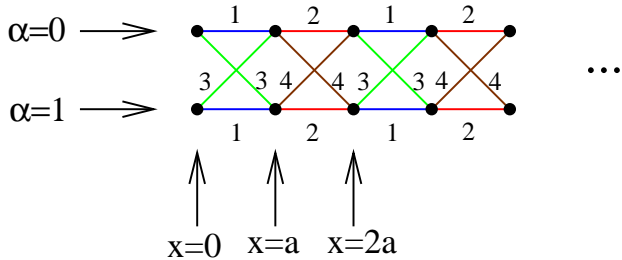}
\caption{\label{coloring} Vertices represent elements of $\{0,1\}
  \times \Omega$ two vertices are connected by an edge if
  $H_{\mathrm{nn}}$ couples these sites. (Different species are never
  coupled by $H_{\mathrm{nn}}$, so the full graph with vertices
  corresponding to elements of $\{1,\ldots,N\} \times \{0,1\} \times
  \Omega$ would consist of $N$ disconnected copies of the graph
  shown.) The edges can be colored with four colors such that each node
  has no more than one incident edge of each color. 
  One can obtain the decomposition
  $H_{\mathrm{nn}} = H_1 + H_2 + H_3 + H_4$ 
  by choosing $H_1$ to be the sum of all interaction terms
  along the edges labeled 1 (which are blue), $H_2$ to be the sum of
  all the interaction terms along edges labeled 2 (which are red), and
  so on.}
\end{center}
\end{figure}

In $H_{\mathrm{nn}}$ there is no coupling between different species,
that is, no products of $\psi_j$ and $\psi_k$ for $j \neq
k$. Thus, we can ignore the index $j$. We now construct a graph whose
vertices correspond to the elements of $\{0,1\} \times \Omega$. We
draw an edge between two vertices if there exists a term in
$H_{\mathrm{nn}}$ with the corresponding support. One sees that this
graph is as shown in Fig.~\ref{coloring}. The graph is
edge-colorable with four colors, and therefore $H_{\mathrm{nn}}$ is
correspondingly decomposable as in \eq{fourcolors} with each of
$H_1,H_2,H_3,H_4$ consisting of a sum of commuting terms. (Because of
the periodic boundary conditions, this works only if $\hat{L}$ is
even, which we assume henceforth.)

The unitary time evolution induced by $H=H_{\mathrm{ss}} +
H_1 + H_2 + H_3 + H_4$ can be approximately decomposed via high-order
Suzuki-Trotter formulae into a sequence of 
\begin{equation}
\label{trottersteps}
n_{\mathrm{S-T}} = O\big((t/a)^{1+o(1)} \hat{L}^{o(1)} \epsilon^{-o(1)} \big)
\end{equation}
time evolutions induced by individual members of $\{H_{\mathrm{ss}},
H_1, H_2, H_3, H_4\}$. The scaling with $t$ follows from
\cite{Suzuki90, Suzuki93}. The scaling with $\hat{L}$ is a consequence
of the spatial locality of $H$ (see \S 4.3 of
\cite{longversion}), that is, the property that only 
nearest-neighbor sites are coupled. 
The scaling with $a$ is a consequence of the fact
that the individual terms in the Hamiltonian each have norm at most of
order $a^{-1}$. This affects the magnitude of the error term in the
Suzuki-Trotter decomposition, which arises from commutators of these
terms.

Each member of $\{H_{\mathrm{ss}}, H_1, H_2, H_3, H_4\}$ is a sum of
$O(\hat{L})$ commuting terms. The time evolution $e^{-i \sum_i
  M_i t}$ induced by 
commuting terms $M_i$
decomposes as $e^{-i \sum_i M_i t} = \prod_i e^{-iM_i t}$. If each
$H_i$ acts on only a constant number of qubits, then the individual
factors $e^{-iH_i t}$ in this product can each be simulated in
$\widetilde{O}(1)$ time, by the Solovay-Kitaev theorem \cite{Kitaev97,
  Dawson_Nielsen}. Thus, including a logarithmic overhead for fermionic
statistics, the cost of implementing $e^{-iJt}$ for any
$J \in \{H_{\mathrm{ss}}, H_1, H_2, H_3, H_4\}$ is $\widetilde{O}(\hat{L})$.
By \eq{trottersteps}, the total cost of time evolution is
$O \big( \left( \frac{tL}{a^2} \right)^{1+o(1)} \epsilon^{-o(1)}
\big)$ quantum gates.


\section{State Preparation and Measurement}

We divide the problem of state preparation into three steps, described
in \sect{freeprep}--\sect{exciting}: 
preparing the free vacuum,
transforming the free vacuum into the interacting vacuum, and exciting
wavepackets on the background of the interacting vacuum.
Two possible measurement procedures are described in 
\sect{measurements} and \sect{sec:charge}.

\subsection{Preparing the Free Vacuum}
\label{freeprep}

Although the free Hamiltonian $H_0 + H_W$ is exactly solvable,
preparing its ground state in the $S_x$ representation on a quantum computer 
is non-trivial. We do so using adiabatic state preparation, as
follows. Let
\begin{equation}
H(s) = \sum_{\bx \in \Omega} a \sum_{j=1}^N \bar{\psi}_j(\bx) \left[
  -si\gamma^1 \frac{\psi_j(\bx + a) - \psi_j(\bx - a)}{2a} + m
  \psi_j(\bx) \right] + s H_W.
\end{equation}
The energy gap of this Hamiltonian is equal to the parameter $m$ for all
$s$. We set this equal to the physical mass of the particles whose
scattering we ultimately wish to simulate.

$H(0)$ is a sum of separate Hamiltonians acting on each lattice
site and each species of particle. Its ground state is therefore the
tensor product of the ground states of the 
four-dimensional Hilbert spaces associated with each pair 
$(\mathbf{x},j) \in \Omega \times \{1,\ldots,N\}$. 
(Specifically, the ground state for a given site is
$\frac{1}{\sqrt{2}} \left( \ket{01} + i\ket{10} \right)$, where
$\ket{b_0 b_1}$ with $b_0,b_1 \in \{0,1\}$ denotes the state
satisfying $a \psi^\dag_{j,0}(\mathbf{x}) \psi_{j,0}(\mathbf{x})
\ket{b_0 b_1} = b_0 \ket{b_0 b_1}$ and
$a \psi^\dag_{j,1}(\mathbf{x}) \psi_{j,1}(\mathbf{x}) \ket{b_0 b_1} =
b_1 \ket{b_0 b_1}$.) The cost of
producing this tensor product of $N \hat{L}$ local states, 
including the cost of fermionic
antisymmetrization via the encoding of \cite{Bravyi_Kitaev},
is $O(N \hat{L} \log (N \hat{L}))$.

After the ground state of $H_0$ has been prepared, the complexity of the
remaining adiabatic state preparation is determined by the
adiabatic theorem \cite{Ruskai, Goldstone}.

\begin{theorem}
\label{adiabaticthm}
Let $H(s)$ be a finite-dimensional twice differentiable Hamiltonian on
$0 \leq s \leq 1$ with a non-degenerate ground state $\ket{\phi_0(s)}$
separated by an energy gap $\gamma(s)$. Let $\ket{\psi(t)}$ be
the state obtained by Schr\"odinger time evolution according to the 
Hamiltonian $H(t/T)$ from the state $\ket{\phi_0(0)}$ at $t = 0$. 
Then, with an appropriate choice of phase for $\ket{\phi_0(t)}$, 
the error $\Delta \equiv \| \Ket{\psi(T)} - \Ket{\phi_0(1)} \|$ satisfies
\begin{equation}
\label{adiabateq}
\Delta \leq
\frac{1}{T} \left[ \frac{1}{\gamma(0)^2} \left\| \frac{\ud H}{\ud s} \right\|_{s=0}
+ \frac{1}{\gamma(1)^2} \left\| \frac{\ud H}{\ud s} \right\|_{s=1}
+ \int_0^1 \ud s \left( \frac{5}{\gamma^3} \left\| \frac{\ud H}{\ud s}
\right\|^2 + \frac{1}{\gamma^2} \left\| \frac{\ud^2 H}{\ud s^2}
\right\| \right) \right]. 
\end{equation}
\end{theorem}

Analyzing the adiabaticity of this process is relatively easy, because
\eq{psij} and \eq{psidagj} diagonalize $H(s)$
(and $\frac{dH}{ds}$) for all $s$. One finds that the eigenvalue gap
of $H(s)$ throughout the adiabatic path $0 \leq s \leq 1$ is always
precisely $m$. Furthermore,
\begin{equation}
\frac{dH}{ds} = \sum_{j=1}^N \sum_{\mathbf{p} \in \Gamma} \frac{1}{L}
E^{(a)}_{\bp}(0) \left( a_j^\dag(\bp) a_j(\bp) + b_j^\dag(\bp) b_j(\bp)
\right).
\end{equation}
Thus,
\begin{equation}
\left\| \frac{dH}{ds} \right\| = 2N \sum_{\bp \in \Gamma} E^{(a)}_{\bp}(0) \,.
\end{equation}
For large $L$, $\sum_{\bp \in \Gamma} \frac{1}{L}$ becomes well
approximated by the integral $\int_0^{2 \pi/a} d \bp$. Thus, using
\eq{eq:Epa}, we obtain 
\begin{eqnarray}
\left\| \frac{dH}{ds} \right\| & \simeq &
2NL \int_0^{2 \pi/a} d \bp E^{(a)}_{\bp}(0) \\
& = & 2NL \int_0^{2 \pi/a} d \bp \sqrt{\frac{4r^2}{a^2} \sin^4 \left(
  \frac{\bp a}{2} \right) + \frac{1}{a^2} \sin^2 (\bp a)} \\
& = & \frac{2NL}{a^2} \eta(r) \,, \label{finalderiv}
\end{eqnarray}
where
\begin{equation}
\eta(r) = \int_0^{2 \pi} d \hat{p} \sqrt{4 r^2 \sin^4 \Big(
  \frac{\hat{p}}{2} \Big) + \sin^2 \left( \hat{p} \right)} \,.
\end{equation}
We can therefore substitute $\frac{d^2 H}{ds^2} = 0$, $\left\|
  \frac{dH}{ds} \right\| = O(L a^{-2})$ and $\gamma = m$ into
  \eq{adiabateq}. Theorem \ref{adiabaticthm} then shows that we can
prepare a state with distance no more than $\epsilon_{\mathrm{prep}}$
from the exact state using
\begin{equation}
T = O \left( \frac{L^2}{a^4 m^3 \epsilon_{\mathrm{prep}}} \right).
\end{equation}
Note that the adiabatic theorem applied here, though convenient because 
of its generality, may not yield a tight upper bound on the run time.

\subsection{Preparing the Interacting Vacuum}
\label{turnon}

\begin{figure}
\begin{center}
\includegraphics[width=0.25\textwidth]{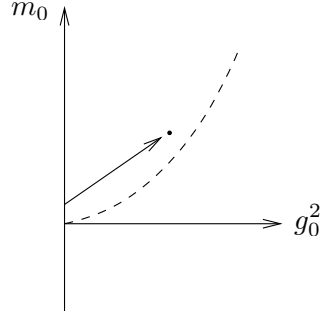}
\caption{\label{paths} Our perturbative calculations of the physical mass
  in the massive Gross-Neveu model indicate a phase diagram with the
  qualitative features illustrated above. The phase above the dashed
  curve is accessible adiabatically from the free theory but the phase below
  is not. The arrow depicts our linear adiabatic path, described in
  \eq{deltam}. Our perturbative analysis shows that the first two
  derivatives of the phase transition curve with respect to $g_0^2$
  are both positive and diverge only as $\mathrm{poly}(\log(m_0 a))$
  in the limit $a \to 0$.
}
\end{center}
\end{figure}

Given the ground state of the free theory, we can prepare the ground
state of the interacting theory by adiabatically varying the parameters
$g_0^2$ and $m_0$ in the massive Gross-Neveu Hamiltonian, starting from
$g_0^2 = 0$. For adiabaticity to be maintained, the physical mass must
not vanish at any point in the adiabatic path. By \sect{massren}, 
the physical mass varies with $g_0^2$ according to
\begin{equation}
\label{two-orders}
m = m_0 - c_1 g_0^2 - c_2 g_0^4 + O(g_0^6) \,,
\end{equation}
where $c_1,c_2>0$ are given by
\begin{eqnarray}
c_1 & = & \frac{m}{2\pi} \log\Big(\frac{1}{ma}\Big) + \cdots \,, \\
c_2 & \simeq & \frac{m}{16\pi^3}\big(9.3 N - 0.07\big)\log^2(ma)  + \cdots\,. 
\label{c2}
\end{eqnarray}
(The coefficients in \eq{c2} were determined numerically.)
The vanishing of the physical mass marks the location of a quantum phase 
transition, which cannot be adiabatically crossed. 
Equation \eq{two-orders} indicates that the phase diagram takes the
schematic form as shown in Fig.~\ref{paths}.

As in \sect{freeprep}, we parametrize our adiabatic state
preparation by a quantity $s$, which increases over time from $0$ to
$1$. In this second adiabatic process, the Hamiltonian is the full
massive Gross-Neveu Hamiltonian with $s$-dependent parameters
$g_0^2(s)$ and $m_0(s)$. We choose $g_0^2(0) = 0$ and $m_0(0)=m$ so
that the initial Hamiltonian of this adiabatic process matches the
final Hamiltonian of the preceding adiabatic step. Thus, the ground
state at $s=0$ is the free vacuum prepared in the previous step of the
algorithm. To keep our analysis simple, we choose a linear adiabatic
path, namely,
\begin{eqnarray}
g_0^2(s) & = & s g_0^2 \,, \nonumber \\
m_0(s) & = & m + s \delta_m \,. \label{linearpath}
\end{eqnarray}
We choose $\delta_m$ so that the physical mass at $s=1$ is
equal to the physical mass at $s=0$. To second order in
$g_0^2$,
\begin{equation}
\label{deltam}
\delta_m = c_1 g_0^2 + c_2 g_0^4 + \cdots \,,
\end{equation}
as illustrated in Fig.~\ref{paths}.

By \eq{linearpath}, $\frac{d^2 H}{ds^2} = 0$ and 
\begin{equation}
\label{deriv}
\frac{dH}{ds} = \sum_{\bx \in \Omega} a \Bigg[
  \delta_m \bar{\psi}_j(\bx) \psi_j(\bx) + \frac{g_0^2}{2} \bigg( \sum_{j=1}^N
\bar{\psi}_j (\bx) \psi_j(\bx) \bigg)^2 \Bigg] \,.
\end{equation}
Furthermore, the minimal eigenvalue gaps occur at $s=0$ and $s=1$ and
are equal to the final physical mass $m$. Thus, to apply Theorem
\ref{adiabaticthm} we need only bound $\left\| \frac{d H}{ds}
\right\|$. 

We can deduce the spectrum of $\frac{dH}{ds}$ by the following
transformation:
\begin{eqnarray}
a_j(\bx) & = & \frac{1}{\sqrt{2}} \big( \psi_{j,0}(\bx) - i
  \psi_{j,1}(\bx) \big) \,,\\
b_j^\dag(\bx) & = & \frac{1}{\sqrt{2}} \big( \psi_{j,0}(\bx) + i
  \psi_{j,1}(\bx) \big) \,.
\end{eqnarray}
This corresponds to
\begin{equation}
\label{localtrans}
\psi_j(\bx) = \frac{1}{\sqrt{2m_0}} \left( a_j(\bx) u(0) + b_j^\dag(\bx)
v(0) \right) \,,
\end{equation}
where $u,v$ are defined in \eq{concrete}. Using \eq{anticanon1} and
\eq{anticanon2}, one can verify that
\begin{eqnarray}
\{ a_j(\bx), a_k^\dag(\by) \} = \{ b_j(\bx), b_k^\dag(\by) \} 
& = & a^{-1} \delta_{j,k} \delta_{\bx, \by} \id \,,\\
\{ a_j(\bx), a_k(\by) \} = \{ b_j(\bx), b_k(\by) \} & = & 0 \,,\\
\{ a_j(\bx), b_k(\by) \} = \{a_j^\dag(\bx), b_k(\by) \} & = & 0 \,.
\end{eqnarray}
Thus, $a_j(\bx),a_j^\dag(\bx),b_j(\bx),b_j^\dag(\bx)$ are
creation and annihilation operators for $2N$ species of fermions
localized on the spatial lattice. By \eq{localtrans},
\begin{equation}
\bar{\psi}_j(\bx) \psi_j(\bx) = a_j^\dag(\bx) a_j(\bx) - b_j(\bx)
b_j^\dag(\bx) \,, 
\end{equation}
from which we obtain
\begin{equation}
\Bigg\| \sum_{j=1}^N \bar{\psi_j}(\bx) \psi_j(\bx) \Bigg\| = 2Na^{-1},
\end{equation}
and hence
\begin{equation}
\left\| \frac{dH}{ds} \right\| = \delta_m 2N\hat{L} + \frac{2
  \hat{L} g_0^2 N^2}{a} \,. 
\end{equation}

By the results of \sect{massren}, we find that $\delta_m =
O(\log^2(ma))$. Hence, recalling that $\hat{L} = L/a$, we obtain
\begin{equation}
\left\| \frac{dH}{ds} \right\| = O \left( \frac{L}{a^2} \right).
\end{equation} 
Therefore, by Theorem \ref{adiabaticthm} the diabatic error is at most
\begin{eqnarray}
\epsilon & = & O \left( \frac{1}{T_{\mathrm{turn-on}}} \frac{ \left\|
      \frac{dH}{ds} \right\|^2}{\gamma^3} \right) \\
& = & O \left( \frac{L^2}{T_{\mathrm{turn-on}} a^4 m^3} \right).
\end{eqnarray}
It thus suffices to choose
\begin{equation}
T_{\mathrm{turn-on}} = O \left( \frac{L^2}{a^4 \epsilon m^3} \right).
\end{equation}

In the above procedure, we choose our adiabatic path so that the
initial and final physical masses equal some user-specified value
$m$. To achieve this, one needs to tune the quantity
$\delta_m$ in accordance with \eq{linearpath} and \eq{deltam}. For
sufficiently weak coupling, the proper choice of $\delta_m$ can be
determined by the perturbative calculations performed in
\sect{massren}. In the strongly coupled case, these perturbative
calculations no longer provide precise guidance as to a choice of
$\delta_m$. Instead, as previously discussed in \cite{longversion},
the adiabatic path can be determined by the quantum
computer. Specifically, one can measure the physical mass at a given
coupling strength $g_0$ by exciting a particle and measuring energy
via phase estimation. This measurement guides the choice of a suitable
adiabatic path to a slightly larger coupling strength, at which
point the mass can be measured again. Iterating this process, one
can reach any coupling strength for which the corresponding vacuum is
in the same quantum phase as the free vacuum.

\subsection{Exciting Wavepackets}
\label{exciting}

After preparing the interacting vacuum, $\ket{\mathrm{vac}}$, we
excite wavepackets by simulating a source that varies sinusoidally in
space and time so as to induce excitations of some particular total energy 
and momentum by resonance. Given the physical rest mass $m$
of the particles, we can choose this energy and momentum so that the
only corresponding state is a single-particle state. (For a given total
momentum, an unbound state of two particles will have greater energy
than the corresponding state of one particle. In the ultrarelativistic
limit, $p \gg m$, this energy difference scales as $m^2/p$.)
In the remainder of this section, we show that, using a source of
spatial extent $l$ and duration $\tau$, one can ensure that
excitations off resonance are suppressed as $\sim \exp \left[
 - \frac{1}{4}  \left( l^2 (\bp-\bp_0)^2 + \tau^2 (E-E_0)^2 \right)
\right]$. Hence, by simulating a process of duration $\tau \sim p/m^2$
and spatial extent $l \sim p/m^2$, one can control the incoming momentum 
and ensure that the probability of obtaining more than one particle is small.

The creation of two incoming particles has only an
$O(\epsilon)$ success probability, which can be compensated for by
repeated attempts. (See the discussion following \eq{w}.) 
The total complexity of preparing two particles is
the cost of simulating the time evolution given in \eq{bigR} a total
of $1/\epsilon$ times. Thus, by the results of \sect{trotter},
the complexity is $\big( \frac{\tau l}{a^2 \epsilon}
\big)^{1+o(1)}$. Thus, since $p \sim a^{-1}$ for fixed
$\epsilon$ and $a \sim \epsilon$ for fixed $p$, the number of quantum
gates $G_{\mathrm{excite}}$ needed to excite the two initial particles
is
\begin{equation}
G_{\mathrm{excite}} \sim \left\{ \begin{array}{ll}
\epsilon^{-3-o(1)}\,, & \textrm{as} \,\,\, \epsilon \to 0 \,, \\
p^{4+o(1)} \,, & \textrm{as} \,\,\, p \to \infty \,.
\end{array} \right.
\end{equation}
Note also that for the initial wavepackets to be well separated, $L$
must be larger than $2l$. Hence, in the high-momentum limit $L \sim
p$, which affects the complexity of other steps of the algorithm.

\subsubsection*{Perturbative Expansion}
\label{dyson}

The resonant excitation can be analyzed with time-dependent perturbation
theory. Let
\begin{equation}
\label{bigR}
R = T \left\{ \exp \left[-i \int_0^\tau \ud t 
\left( H+ \lambda W(t) \right) \right] \right\} \,,
\end{equation}
where $T\{ \cdot \}$ denotes the time-ordered product, 
$H$ is given by \eq{h},
\begin{equation}
\label{Wt}
W(t) = \int \ud \bx \left( f(t,\bx)\psi_{i,\alpha}(\bx) +
  f^*(t,\bx)\psi_{i,\alpha}^\dag(\bx) \right),
\end{equation}
$i$ and $\alpha$ are chosen
according to the desired type of particle, and $f(t,\bx)$ is an 
oscillatory function whose form we optimize in the next
subsection. The end product of the excitation process is $R
\ket{\mathrm{vac}}$. One can expand this quantity using the Dyson
series, as follows:
\begin{equation}
\label{Dyson}
R = \id - i \lambda \int_0^\tau \ud t_1 W_I(t_1) + (-i \lambda)^2
\int_0^\tau \ud t_1 \int_0^{t_1} \ud t_2 W_I(t_1) W_I(t_2) + \cdots \,,
\end{equation}
where 
\begin{equation}
\label{WI}
W_I(t) = e^{iHt} W(t) e^{-iHt}
\end{equation}
and the $n\th$-order term in $\lambda$ is
\begin{equation}
(-i \lambda)^n \int_0^\tau \ud t_1 \ldots \int_0^{t_{n-1}} d t_n
W_I(t_1) \ldots W_I(t_n) \,.
\end{equation}
The total contribution from orders $\lambda^2$ and higher is bounded by
\begin{eqnarray}
\left\| \sum_{n=2}^\infty (-i \lambda)^n \int_0^\tau \ud t_1
  \ldots
\int_0^{t_{n-1}} \ud t_n W_I(t_1) \ldots W_I(t_n) \right\|
& \leq & \sum_{n=2}^\infty \frac{\lambda^n \tau^n}{n!} w^n \\
& = & \exp[\lambda \tau w] - 1 - \lambda \tau w \,,
\end{eqnarray}
where
\begin{equation}
\label{w}
w = \max_{0 \leq t \leq \tau} \left\| W(t) \right\|.
\end{equation}

From the above analysis, one sees that the Dyson series converges
rapidly. The single-particle excitation amplitude is of order
$\lambda$, and the dominant error, other than non-excitation, is the
two-particle excitation amplitude, which is of order
$\lambda^2$. Setting the two-particle excitation probability to
$\epsilon$, one obtains a single-particle excitation with probability
$p_1 \sim \sqrt{\epsilon}$, and non-excitation with probability on the order of
$1-\sqrt{\epsilon}$. In a standard scattering simulation, one wishes
to prepare as an initial state single-particle excitations at two
spatially separated locations. The fraction of simulations in which
this is achieved (rather than one or both particles failing to be produced) 
is thus of order $p_1^2 \sim \epsilon$. One
can detect such instances and compensate by repeating the
simulation $O(1/p_1^2)$ times and postselecting the instances in
which both particles were produced.

Next, we consider the first-order excitation amplitude in more
detail. Let $\ket{E,\bp}$ be any state with total momentum $\bp$ and
energy $E$ above the vacuum energy, so that  $P \ket{E,\bp} = \bp
\ket{E,\bp}$ and $H \ket{E,\bp} = E \ket{E,\bp}$, where $P$ is the total
momentum operator. (Here, we rely on the fact that $[H,P] = 0$.) Then,
to first order in $\lambda$, by \eq{Dyson} and \eq{WI},
\begin{eqnarray}
\bra{E,\bp} R \ket{\mathrm{vac}} 
& \simeq & - i \lambda \int_0^\tau \ud t \bra{E,\bp} W_I(t)
\ket{\mathrm{vac}} \\
& = & - i \lambda \int_0^\tau \ud t \, e^{-iEt} \bra{E,\bp} W(t)
\ket{\mathrm{vac}} \,.
\end{eqnarray}
Recalling that the momentum operator is the generator of spatial
translations, one has $\psi_{i,\alpha}(\bx) = e^{iP\bx} \psi_{i,\alpha}(0)
e^{-iP\bx}$. Thus, to first order in $\lambda$,
\begin{equation}
\bra{E,\bp} R \ket{\mathrm{vac}} 
\simeq - i \lambda \int_0^\tau \ud t \int
\ud \bx e^{-i(Et+\bp\bx)} \left[ f(t,\bx) \bra{E,\bp} \psi_{i,\alpha}(0)
  \ket{\mathrm{vac}} + f^*(t,\bx) \bra{E,\bp} \psi^\dag_{i,\alpha}(0)
  \ket{\mathrm{vac}} \right] \,.\\
\end{equation}
(Here we have used $P \ket{\mathrm{vac}} = 0$.) Defining $f(t,\bx) = 0$
for $t \notin [0,\tau]$, we can extend the time integration to
infinity and express $\bra{E,\bp} R \ket{\mathrm{vac}}$ in terms of
$\tilde{f}$, the Fourier transform of $f$. For our choice of $f$,
given in the next subsection, $\tilde{f}$ is real, and therefore
\begin{equation}
\label{amp}
\bra{E,\bp} R \ket{\mathrm{vac}} =  - i \lambda 
\left[ \tilde{f}(E,\bp) \bra{E,\bp} \psi_{i,\alpha}(0) \ket{\mathrm{vac}}
+ \tilde{f}(-E,-\bp) \bra{E,\bp} \psi_{i,\alpha}^\dag(0)
\ket{\mathrm{vac}} \right] + O(\lambda^2).
\end{equation}

\subsubsection*{Wavepacket Shaping}
\label{shaping}

We now show that a Gaussian wavepacket is a good choice for $f(t,\bx)$.
Specifically, for chosen constants $\alpha, \beta > 0$, let
\begin{equation}
f(t,\bx) = \left\{ \begin{array}{cl}
\eta \exp \left[-(\alpha t)^2 - (\beta \bx)^2 - i E_0 t + i \bp_0 \bx \right] \,, &
-\tau/2 \leq t \leq \tau/2, -l/2 \leq \bx \leq l/2 \,, \\
0 \,, & \textrm{otherwise} \,.
\end{array} \right.
\end{equation}
(For convenience, we have shifted the origin of the coordinate
system.) Here $\eta$ is a normalization factor\footnote{It is 
  reasonable to choose $\eta$ so that $\int_0^{\tau} dt W_I(t)
  \ket{\mathrm{vac}}$ is a normalized state. In the ultrarelativistic
  limit this implies that $\eta \sim \left( \alpha^2
    \beta^4 + \alpha^4 \beta^2 \right)^{1/4}$.} with mass dimension $3/2$. 
With this choice of $f$,
\begin{equation}
\label{1peak}
\tilde{f}(E,\bp) =  \eta q_{\beta,l}(\bp-\bp_0) q_{\alpha,\tau}(E-E_0) \,,
\end{equation}
where
\begin{equation}
q_{\rho,r}(d) = \int_{-r/2}^{r/2} \ud \bx \, e^{id\bx-(\rho \bx)^2}.
\end{equation}
In the limit $r \to \infty$, the function $q_{\rho,r}(d)$ converges to a
Gaussian peak of width $\sim 1/\rho$. Since $E$ must be positive, the
$\tilde{f}(-E,-\bp) \bra{E,\bp} \psi^\dag_{i,\alpha}(0)
\ket{\mathrm{vac}}$ term in \eq{amp} is exponentially small. Hence,
one obtains
\begin{equation}
\label{amp2}
\bra{E,\bp} R \ket{\mathrm{vac}} \simeq  - i \lambda 
\tilde{f}(E,\bp) \bra{E,\bp} \psi_{i,\alpha}(0) \ket{\mathrm{vac}}.
\end{equation}
for $E \gg 1/\tau$ and $\lambda \ll 1$. By \eq{amp2} and \eq{psij}, one
sees that $R \ket{\mathrm{vac}}$ is a antifermion wavepacket with
momentum centered around $\bp$. To create a fermion, one interchanges
$\psi$ and $\psi^\dag$ in \eq{Wt}.

Using the asymptotics of error functions, we can furthermore bound the
contributions due to $r$ being finite. One finds that
\begin{equation}
\big|q_{\rho,r}(d) - q_{\rho,\infty}(d)\big|  \leq  \frac{2}{r \rho^2}
e^{-(\rho r)^2/4}.
\end{equation}

\subsection{Measuring Number Operators}
\label{measurements}

Recall from \sect{rep} that the free theory ($g_0^2 = 0$) 
is exactly solvable, with the number operators
$L^{-1} a_j^\dag(\mathbf{p}) a_j(\mathbf{p})$ counting fermions of species
$j$ in momentum-mode $\mathbf{p}$ and $L^{-1} b_j^\dag(\mathbf{p})
b_j(\mathbf{p})$ similarly counting antifermions. 
Thus, as one possible set of measurements to perform on the final state of 
the simulation, we propose, as in \cite{longversion}, adiabatically returning 
to the free theory and then measuring number operators via the 
phase-estimation algorithm. 
We analyze this measurement procedure in this section. 
An alternative set of measurements that is more suitable when bound states
are present is analyzed in \sect{sec:charge}.

The adiabatic return to the free theory is performed in the presence
of particle wavepackets, so the state being adiabatically
transformed is not an energy eigenstate. Different energy eigenstates
in the superposition will acquire different dynamical phases
during the adiabatic process and thus, in physical terms, the simulated
particles will propagate. Such propagation is undesirable because we do 
not want any scattering to occur while the interaction is being turned off. 

Hence, we apply the same technique proposed in \cite{longversion} to
suppress particle propagation: we interleave (simulated) backwards
time evolutions governed by time-independent Hamiltonians into the
adiabatic process. By an analysis similar to that performed in
\cite{longversion}, one finds that, to ensure that a particle
propagates no further than a distance $\mathcal{D}$, it suffices to use
\begin{equation}
J = \widetilde{O} \left( \frac{\sqrt{\tau}}{p \mathcal{D}} \right)
\end{equation}
backwards evolutions, where $\tau$ is the duration of the original
adiabatic process and $p$ is the momentum of the particle. Further,
one finds that the total probability of diabatically exciting one or
more particles is\footnote{This result is based on the
  adiabatic criterion of \cite{Messiah} which
  appears to be applicable \cite{longversion} to our Hamiltonian 
  although it may not apply to all Hamiltonians.}
\begin{equation}
P_{\mathrm{diabatic}} = O\left( \frac{J^2 L p^2}{\tau^2} \right).
\end{equation}
Hence, setting $\mathcal{D}$ to a constant
$P_{\mathrm{diabatic}}$ to  $\epsilon$, one obtains
\begin{equation}
\tau = \widetilde{O} \left( \frac{L}{\epsilon} \right). 
\end{equation}
A process of this duration can
be implemented with (\sect{trotter})
\begin{equation}
G_{\mathrm{turn-off}} = O \left( \left(
    \frac{L^2}{a \epsilon} \right)^{1+o(1)} \right)
\end{equation}
quantum gates.

The phase-estimation algorithm \cite{Kitaev95} enables one to
measure in the eigenbasis of $L^{-1} a_j^\dag(\mathbf{p}) a_j(\mathbf{p})$,
provided one can efficiently implement $e^{-i L^{-1} a_j^\dag(\mathbf{p})
  a_j(\mathbf{p}) t}$ for various $t$ using quantum circuits. By
\eq{adef} and \eq{bdef}, one sees that the problems of simulating 
$e^{-i L^{-1} a_j^\dag(\mathbf{p}) a_j(\mathbf{p}) t}$ and its
antifermionic counterpart are largely similar to the problem of
simulating the time evolution $e^{-iHt}$, which was analyzed in detail
in \sect{trotter}. However, these number operators are spatially
nonlocal, which means that the methods of \sect{trotter} do not
perform well as a function of $\hat{L}$. Instead, it is more
efficient to use recent techniques from \cite{BCCKS}. 

In \cite{BCCKS}, a method is described for simulating sparse
Hamiltonians in which the matrix elements are given by an oracle. As
discussed on pg. 2 of \cite{BCCKS}, in the case where the sparse
Hamiltonian consists of a sum of $d$ terms each acting on $O(1)$ qubits,
the number of oracle queries and non-oracle-related quantum gates both
scale as $O(d)$. A number operator for a momentum mode consists of
$O(\hat{L}^2)$ terms, acting between all pairs of spatial lattice sites.
Thus, if one ignored the fermionic statistics, the number of non-oracle-related
gates needed to simulate the time-evolution induced by a number operator
would be $O(\hat{L}^2 n) = O(\hat{L}^3)$. The number of gates needed to
implement one oracle query to the sparse matrix defined by the number
operator would be $O(n)$, and number of quantum gates needed to implement
all of the oracle queries would be $O(\hat{L}^3)$. Using the Bravyi-Kitaev
encoding for fermionic statistics adds a logarithmic factor to the
complexity. Measuring all $2N\hat{L}$ of the number operators thus has total
complexity $\widetilde{O}(\hat{L}^4) = \widetilde{O}(L^4/a^4)$.

\subsection{Measuring Local Charge}
\label{sec:charge}

In previous work \cite{longversion}, we proposed measuring local
energy observables as an alternative to returning to the free theory
and measuring number operators. This procedure has the advantage that
it can detect bound states. It has the disadvantage that the local
energy observables have ultraviolet-divergent vacuum fluctuations 
that represent a noise background above which particle excitations must be
discerned. 
In this paper, we instead measure simpler local observables, namely 
charges, whose vacuum fluctuations are less difficult to
control. These observables can thus detect charged bound states,
although they are blind to neutral ones.

From the equation of motion of the massive Gross-Neveu model, one finds 
that for each $j \in \{1,2,\ldots,N\}$ the
quantity
\begin{equation}
J^\mu_j(x) = \bar{\psi}_j(x) \gamma^\mu \psi_j(x)
\end{equation}
obeys
\begin{equation}
\partial_\mu J_j^\mu = 0.
\end{equation}
Hence,
\begin{equation}
\widetilde{Q}_j \equiv \sum_{\mathbf{x}} J_j^0(\mathbf{x}) = \sum_{\mathbf{x}}
\bar{\psi}_j(\mathbf{x}) \gamma^0 \psi_j(\mathbf{x})
\end{equation}
is a conserved charge. Note that, for any 
$b,c \in \mathbb{R}$, $Q_j = b \widetilde{Q}_j + c$ is also
conserved. We can calibrate the charge observable by demanding that
the vacuum have zero charge and that particle creation change the
charge by $\pm 1$. One satisfies these criteria with the following
definition:
\begin{equation}
Q_j = \sum_{\mathbf{x} \in \Omega} a \bar{\psi}_j(\mathbf{x}) \gamma^0
\psi_j(\mathbf{x}) - \hat{L} \id \,.
\end{equation}
By (\ref{psij}), (\ref{psidagj}), and (\ref{anticanonp2}), one finds that
\begin{equation}
Q_j = \frac{1}{L} \sum_{\mathbf{p} \in \Gamma} \left(
  a_j^\dag(\mathbf{p}) a_j(\mathbf{p}) - b_j^\dag(\mathbf{p})
  b_j(\mathbf{p})\right).
\end{equation}
For any envelope function $f:\Omega \to [0,1]$, one can similarly
define
\begin{equation}
Q_j^{(f)} =  \sum_{\mathbf{x} \in \Omega} f(\mathbf{x}) \left( a
  \bar{\psi}_j(\mathbf{x}) \gamma^0 \psi_j(\mathbf{x}) - \id \right).
\end{equation}
If $f$ has support only in some region $R \subset \Omega$, then
$Q_j^{(f)}$ can be thought of as an observable for the charge in that
region. 

The most obvious choice of $f$ is a square function that is
equal to one inside $R$ and zero elsewhere. However, 
a better signal-to-noise ratio can be obtained by choosing $f$ to
decay from one to zero more smoothly at the boundary of $R$.
Specifically, calculations (in Appendix \ref{fluctuations}) show that, 
when $f$ is chosen to be a
Gaussian of width $R$, the variance of the observable $Q_j^{(f)}$ in
the vacuum state is $O(1/mR)$, independent of the lattice spacing $a$. 
Hence the noise background above which particle excitations are to be
detected is nondivergent in $a$ and can be brought to an arbitrarily
low level at the cost of increasing the detector size. 
In practice, one will use a truncated Gaussian, replacing the
exponentially small tails with zero at distances greater than some constant
multiple of $R$. This modified $f$ then has support on a region of
size $O(R)$, but the corresponding operator is exponentially close to
the Gaussian case treated by our analysis.

$Q_j^{(f)}$ has eigenvalues with $O(1)$ separations. Thus, measuring
$Q_j^{(f)}$ by phase estimation entails simulating the unitary
transformation $\exp\big[i Q_j^{(f)} t\big]$ for $t$ of order one. Because
$Q_j^{(f)}$ is the sum of local terms, these unitary
transformations can be implemented by techniques similar to those
in \sect{trotter} with complexity $O(a^{-1-o(1)}
\epsilon^{-o(1)})$.


\section{Some Field-Theoretical Aspects}

This section describes some quantum field-theoretical calculations:
analysis of the effect of discretizing the spatial dimension of the 
massive Gross-Neveu model, and the perturbative renormalization of 
the mass in the discretized theory.

In our complexity analysis (\sect{sec:complexity}), our criterion for
choosing the lattice spacing $a$ is that the scattering cross sections 
for processes at a momentum scale $p$ in the discretized theory should 
differ from their continuum values by at most a factor of $(1+\epsilon)$. 
The results of \sect{EFT} show that one can satisfy this criterion 
by choosing $a \sim \epsilon/p$.
This choice then affects the overall scaling of the algorithm in the
large-momentum and high-precision limits. As one would expect,
higher energies and greater precision require a smaller lattice spacing
and thus a larger number of lattice sites (for fixed $L$).  
Consequently, the number of quantum gates needed to simulate
time evolutions via Suzuki-Trotter formulae is larger.

In \sect{massren}, we perturbatively calculate the relationship
between the bare mass $m_0$, which is a parameter in the
lattice Hamiltonian (see \eq{h} and \eq{h0}), and the physical mass
$m$ of the particles in the theory. We need to know the behavior of
$m$ in order to design and analyze the procedure for preparing the
interacting vacuum (\sect{turnon}).  In particular, a suitable
adiabatic path must maintain a non-zero mass, the magnitude of which
affects the algorithmic complexity, as indicated by the adiabatic
theorem.



\subsection{Effects of Non-zero Lattice Spacing}
\label{EFT}

The effects of a non-zero lattice spacing can be analyzed via effective
field theory. The discretized Lagrangian can be thought of as the leading 
contribution to an effective field theory, neglected terms of which 
correspond to discretization errors. Hence, the scaling of the error 
with the lattice spacing is given by the scaling of the coefficients of 
those terms. 

The symmetries of the continuum theory restrict the possible operators 
in the effective field theory.
Consider the discrete transformations parity (denoted $P$), 
time reversal ($T$), and charge conjugation ($C$).
Parity changes the handedness of space and hence reverses the momentum.
Thus,
\begin{equation}
P a(\bp) P = a(-\bp) \,,\qquad
P b(\bp) P = -b(-\bp) \,.
   \label{eq:P}
\end{equation}
Using \eq{eq:psi} and \eq{eq:P}, we then obtain
\begin{equation}
P \psi(t,\bx) P = \gamma^0\psi(t,-\bx) \,,\qquad
P \bar{\psi}(t,\bx) P = \bar{\psi}(t,-\bx)\gamma^0 \,.
\end{equation}
Likewise, 
\begin{equation}
T a(\bp) T = a(-\bp) \,,\qquad
T b(\bp) T = -b(-\bp) \,.
\end{equation}
It turns out that time reversal needs to be an antilinear operator.
Then
\begin{equation}
T \psi(t,\bx) T = \gamma^1\psi(-t,\bx) \,,\qquad
T \bar{\psi}(t,\bx) T = -\bar{\psi}(-t,\bx)\gamma^1 \,.
\end{equation}
Finally, charge conjugation interchanges particles and antiparticles.
Thus,
\begin{equation}
C a(\bp) C = b(\bp) \,,\qquad
C b(\bp) C = a(\bp) \,,
\end{equation}
and
\begin{equation}
C \psi(t,\bx) C = \psi^*(t,\bx) \,,\qquad
C \bar{\psi}(t,\bx) C = \psi^T(t,\bx)\gamma^0 \,.
\end{equation}
One can verify that the Lagrangian (\ref{eq:MGN}) is invariant under
each of the transformations $P$, $T$ and $C$.

Now consider the operator $\psi^\dagger \mathbb{M} \psi$, where
$\mathbb{M}$ is Hermitian. 
Invariance under $P$, $T$ and $C$ requires
\begin{eqnarray}
\mathbb{M} & = & \gamma^0 \mathbb{M} \gamma^0 \,, \\
\mathbb{M} & = & -\gamma^1 \mathbb{M}^* \gamma^1 \,, \\
\mathbb{M} & = & - \mathbb{M}^T \,. 
\end{eqnarray}
These conditions imply that
\begin{equation}
\mathbb{M} = c \gamma^0\,, \,\, c \in \mathbb{R} \,.
\end{equation}
Likewise, for $i\psi^\dagger \mathbb{M} \partial_\mu \psi$,
where $\mathbb{M}$ is Hermitian, $P$, $T$ and $C$ invariance requires
\begin{eqnarray}
\mathbb{M} & = & (-1)^\mu \gamma^0 \mathbb{M} \gamma^0 \,, \\
\mathbb{M} & = & -(-1)^\mu \gamma^1 \mathbb{M}^* \gamma^1 \,, \\
\mathbb{M} & = & \mathbb{M}^T \,. 
\end{eqnarray}
These conditions imply that, for $\mu=0$,
\begin{equation}
\mathbb{M} = c \id = c (\gamma^0)^2 \,, \,\, c \in \mathbb{R} \,,
\end{equation}
while, for $\mu=1$,
\begin{equation}
\mathbb{M} = c \gamma^5 = -c\gamma^0\gamma^1 \,, \,\, c \in \mathbb{R} \,.
\end{equation}
Thus, the only $P$-, $T$- and $C$-invariant bilinears of Dirac fields are
$\bar{\psi}\psi$ and $i\bar{\psi}\gamma^\mu\partial_\mu \psi$
($\mu=0$ or $1$).

Now consider four-fermion operators, namely, products of two bilinears.
The set $\{\id,\sigma^i\}$ forms a complete basis, elements of which
satisfy the identity
\begin{eqnarray}
\delta_{\alpha\beta} \delta_{\gamma\delta} 
& = & \frac{1}{2}\big(\delta_{\alpha\delta} \delta_{\gamma\beta} 
+ \sum_{i=1}^{3} \sigma^i_{\alpha\delta} \sigma^i_{\gamma\beta}\big)
\,.
\end{eqnarray}
For 
$\gamma^0 = \sigma^2$, $\gamma^1 = -i\sigma^1$, $\gamma^5 = \sigma^3$,
this is equivalent to
\begin{eqnarray}
\delta_{\alpha\beta} \delta_{\gamma\delta} 
& = & \frac{1}{2}(\delta_{\alpha\delta} \delta_{\gamma\beta} 
+ (\gamma^\mu)_{\alpha\delta} (\gamma_\mu)_{\gamma\beta}
+ (\gamma^5)_{\alpha\delta} (\gamma^5)_{\gamma\beta})
\,.
   \label{eq:Fierz0}
\end{eqnarray}
Equation~(\ref{eq:Fierz0}) can be used to obtain Fierz identities.
For example,
\begin{eqnarray}
\bar{\psi}_i\psi_j \bar{\psi}_j\psi_i 
& = & (\bar{\psi}_i)_\alpha(\psi_j)_\beta 
(\bar{\psi}_j)_\gamma(\psi_i)_\delta \delta_{\alpha\beta}\delta_{\gamma\delta} 
\nonumber \\
& = & -\frac{1}{2} \big(\bar{\psi}_i\psi_i \bar{\psi}_j\psi_j
+ \bar{\psi}_i\gamma^\mu\psi_i \bar{\psi}_j\gamma_\mu\psi_j
+ \bar{\psi}_i\gamma^5\psi_i \bar{\psi}_j\gamma^5\psi_j \big)\,,
\end{eqnarray}
where the minus sign comes from fermion anticommutation.
Thus, any operator of the form 
$\bar{\psi}_i\tilde\Gamma_1\psi_j \bar{\psi}_j\tilde\Gamma_2\psi_i$ 
can be rewritten as a sum of operators of the form 
$\bar{\psi}_i\Gamma_1\psi_i \bar{\psi}_j\Gamma_2\psi_j$,
with $\Gamma_{1,2} \in \{\id, \gamma^\mu, \gamma^5 \}$.

If $\Gamma_1 \neq \Gamma_2$, then 
$\bar{\psi}_i\Gamma_1\psi_i \bar{\psi}_j\Gamma_2\psi_j$ will violate
at least one of the discrete symmetries.
Furthermore, the $O(N)$ symmetry\footnote{In fact, the massive Gross-Neveu 
model has an $O(2N)$ symmetry. 
} 
associated with the $N$ fermion species restricts 
the allowed form of operators to functions of 
$\sum_{i=1}^N \bar{\psi}_i \Gamma \psi_i$.
For $i\neq j$, $\bar{\psi}_i\gamma^5\psi_i \bar{\psi}_j\gamma^5\psi_j$
is ruled out by invariance under $P$ (or $C$) of any single
field $\psi_i$, and thus 
$\big(\sum_{i=1}^N \bar{\psi}_i \gamma^5 \psi_i\big)^2$ is ruled out. 
Likewise, 
$\bar{\psi}_i\gamma^\mu\psi_i \bar{\psi}_j\gamma_\mu\psi_j$ ($i\neq j$)
and consequently
$\big(\sum_{i=1}^N \bar{\psi}_i \gamma^5 \psi_i\big)^2$ 
are ruled out.

We conclude that the only four-fermion operator (without derivatives) 
in the effective field theory is 
$(\sum_{i=1}^N \bar{\psi}_i \psi_i)^2$.

Each extra derivative or factor of $\bar{\psi}\Gamma\psi$ in an operator 
will increase its mass dimension by one; correspondingly, it will be 
suppressed by an extra power of $a$.
We therefore conclude that no new unsuppressed operators are
induced in the effective field theory.
The spatial derivative in the continuum theory is approximated by a 
difference operator, with an error of order $a$, and the Wilson term is
also formally of order $a$.
Spatial discretization errors are hence of order $a$.

\subsection{Mass Renormalization}
\label{massren}

In this subsection, we calculate the renormalized (or physical) mass of the
discretized theory, using second-order perturbation theory. 
A convenient way to obtain a suitable expression is to use a partially 
renormalized form of perturbation theory (as was done in \cite{longversion}), 
in which one uses the bare coupling but the renormalized mass. 

To perform perturbative calculations, we need the Feynman rules
for the discretized theory.
The propagator is
\begin{equation}
\begin{array}{l}
\includegraphics[width=0.6in]{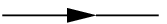} 
\end{array} 
= 
\frac{\gamma^\mu\tilde{p}_\mu+\widetilde{m}(p)}{\tilde{p}^2-\widetilde{m}(p)^2} 
\,,
\end{equation}
where
\begin{equation}
\tilde{p}^\mu = \left( p^0,\frac{1}{a}\sin(a p^1) \right) ,\qquad 
\widetilde{m}(p) = m + \frac{2r}{a} \sin^2\left(\frac{a p^1}{2}\right)
\,.
\end{equation}
For convenience, we use the standard technique of introducing an auxiliary
field $\sigma$ and rewrite the Lagrangian as
\begin{equation}
{\cal L} = {\cal L}_0 + {\cal L}_\sigma \,,
\end{equation}
where ${\cal L}_0$ is the discretized free Lagrangian and
\begin{equation} 
 {\cal L}_\sigma = -\frac{1}{2} \sigma^2 - g \sigma \bar{\psi}_j \psi_j
\,.
\end{equation}
The corresponding Feynman rules are
\begin{equation} \label{y}
\begin{array}{l}
\includegraphics[width=0.6in]{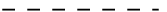} 
\end{array} 
 = -i \,,\qquad
\begin{array}{l}
\includegraphics[width=0.6in]{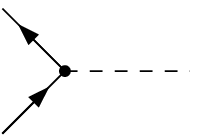} 
\end{array} 
 = -ig \,.
\end{equation}

At one-loop order,
\begin{eqnarray}
-i M(p)
& = & 
\begin{array}{l} \includegraphics[width=0.6in]{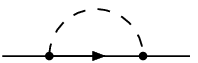} 
\end{array} 
+
\begin{array}{l} \includegraphics[width=0.6in]{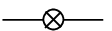} 
\end{array} 
\,,
  \label{eq:diags}
\end{eqnarray}
where the second diagram is the counterterm.

The first diagram gives
\begin{eqnarray}
\begin{array}{l}
\includegraphics[width=0.6in]{dia1b.eps} 
\end{array} 
& = & -g_0^2 \int_{-\infty}^{\infty} \frac{dk^0}{2\pi} 
\int_{-\pi/a}^{\pi/a} \frac{dk^1}{2\pi}
\frac{\gamma^\mu\tilde{k}_\mu+\widetilde{m}(k)}{\tilde{k}^2-\widetilde{m}(k)^2} 
\\
& = & \frac{ig_0^2}{4\pi a} \int_{-\pi}^{\pi} dk^1
\frac{ma + 2r \sin^2\big(\frac{k^1}{2}\big)}{
\sqrt{ \sin^2 k^1 + \big( ma + 2r \sin^2\big(\frac{k^1}{2}\big) \big)^2}
}
\,.
\label{eq:m1}
\end{eqnarray}
The term in \eq{eq:m1} proportional to $r$ scales as $1/a$ and gives the 
mass correction to the doubler (spurious fermion). The term proportional to 
$m$ gives the following:
\begin{equation}
m_0 = m - \frac{g_0^2 m}{2\pi} \log(ma) + \cdots \,.
\end{equation}

At two-loop order, the 1PI amplitude has the additional contributions
\begin{eqnarray}
\begin{array}{l} \includegraphics[width=0.6in]{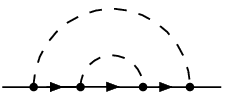}
\end{array}
+
\begin{array}{l} \includegraphics[width=0.6in]{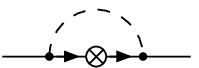} 
\end{array} 
+
\begin{array}{l} \includegraphics[width=0.6in]{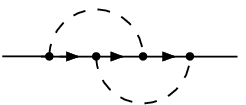}
\end{array}
+
\begin{array}{l} \includegraphics[width=0.6in]{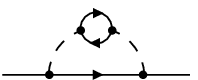} 
\end{array} 
\,.
\nonumber
  \label{eq:diags2}
\end{eqnarray}
The renormalization condition satisfied at first order implies that
the first two diagrams cancel.



The last two diagrams give
\begin{equation}
\begin{array}{l}
\includegraphics[width=0.6in]{dia2b.eps} 
\end{array} 
 = -\frac{ig_0^4}{16\pi^3}\left(m I_1^{(a)} + \frac{1}{a} I_1^{(b)}\right) 
\end{equation}
and
\begin{equation}
\begin{array}{l}
\includegraphics[width=0.6in]{dubub2.eps} 
\end{array} 
 = \,\frac{ig_0^4 N}{16\pi^3}\left(m I_2^{(a)} + \frac{1}{a} I_2^{(b)}\right) 
\,,
\end{equation}
where $I_1^{(a)}$, $I_1^{(b)}$, $I_2^{(a)}$ and $I_2^{(b)}$ are given
in Appendix \ref{integrals}. Numerical evaluation of these integrals
reveals the forms
\begin{eqnarray}
I_{i}^{(b)} & = & c^{(b1)} - c^{(b2)} \, m a + \cdots \,, \\
I_{i}^{(a)} & = & c_{i}^{(a1)} \log^2(ma) - c_{i}^{(a2)} \log(ma)
+ c_{i}^{(a3)} + \cdots \,, 
\end{eqnarray}
with $c_i^{(j)} > 0$.
We thus obtain
\begin{equation}
m = m^{(1)} - \frac{g_0^4m^{(1)}}{16\pi^3}\big(N c_2^{(a1)}-c_1^{(a1)}\big) 
\log^2(m^{(1)}a)
+ \cdots \,,
\end{equation}
where $m^{(1)}$ denotes the physical mass at one-loop
order. 

\bigskip
\bigskip

\noindent \textbf{Acknowledgments:}
We thank William George for help with numerical calculations. This
work was supported by NSF grant PHY-0803371, DOE grant
DE-FG03-92-ER40701, and NSA/ARO grant W911NF-09-1-0442. IQC and
Perimeter Institute are supported in part by the Government of Canada
through Industry Canada and by the Province of Ontario through the
Ministry of Research and Innovation. The Institute for Quantum
Information and Matter (IQIM) is an NSF physics Frontiers Center with
support from the Gordon and Betty Moore Foundation. S.J. and K.L. are
grateful for the hospitality of the IQIM (formerly IQI), Caltech,
during parts of this work. Portions of this work are a contribution of
NIST, an agency of the US Government, and are not subject to US
copyright.


\appendix

\newpage
\section{Variance of Local Charge}
\label{fluctuations}

Consider $\widetilde{Q}_j^{(f)}$ in the continuum limit:
\begin{equation}
\widetilde{Q}_j^{(f)} = \int dx J_j^{(0)}(\mathbf{x}) f(\mathbf{x}) \,.
\end{equation}
We wish to compute its variance, which is given by
\begin{eqnarray}
\label{eq:fluct}
\Big\langle \big(\widetilde{Q}_j^{(f)} - \big\langle \widetilde{Q}_j^{(f)} 
\big\rangle \big)^2 \Big\rangle 
& = & 
\int dx dy f(\mathbf{x}) f(\mathbf{y})
\big( \langle J_j^{(0)}(\mathbf{x}) J_j^{(0)}(\mathbf{y})\rangle 
- \langle J_j^{(0)}(\mathbf{x}) \rangle \langle J_j^{(0)}(\mathbf{y}) \rangle
\big) \\
& = & 
\int \frac{d^2k}{(2\pi)^2} |\widetilde{f}(k)|^2 \widetilde{G}_c(k) \,,
\label{eq:fluct2}
\end{eqnarray}
where $G_c$ is the connected Green's function.
By standard quantum field-theoretical methods, we obtain
\begin{eqnarray}
\label{eq:Gc}
\widetilde{G}_c(k^0,k^1) & = & 2i \int_0^1 dx \int \frac{d^2p_E}{(2\pi)^2}
\frac{m^2 - x (1-x)((k^0)^2 + (k^1)^2)}{[p_E^2 +m^2 
- x(1-x)((k^0)^2-(k^1)^2)]^2}
\,, \\
& = & \frac{i}{2\pi} \int_0^1 dx \,
\frac{m^2 - x (1-x)((k^0)^2 + (k^1)^2)}{m^2 - x(1-x)((k^0)^2-(k^1)^2)} \,.
\label{eq:Gc2}
\end{eqnarray}
Substituting $\widetilde{G}_c$ into \eq{eq:fluct2} and using an ultraviolet 
regulator, we obtain
\begin{equation}
\Big\langle \big(\widetilde{Q}_j^{(f)} - \big\langle \widetilde{Q}_j^{(f)} 
\big\rangle \big)^2 \Big\rangle 
 = \frac{1}{(2\pi)^2} \int dk_1 |\tilde{f}(k^1)|^2 \int_0^1 dx \,\,
\frac{(k^1)^2}{\sqrt{(k^1)^2 + \frac{m^2}{x(1-x)}}} \,.
\end{equation}

For the square window function $f(x) = \theta(R/2 - |x|)$, the
charge fluctuation diverges, but only logarithmically, with 
\begin{equation}
\Big\langle \big(\widetilde{Q}_j^{(f)} - \big\langle \widetilde{Q}_j^{(f)} 
\big\rangle \big)^2 \Big\rangle 
 \leq \frac{2}{\pi^2} \big(\log(2\pi)-1-\log(ma)\big) + O((ma)^2) \,.
\end{equation}
For $f(x) = \exp(-x^2/R^2)$,
\begin{equation}
\Big\langle \big(\widetilde{Q}_j^{(f)} - \big\langle \widetilde{Q}_j^{(f)} 
\big\rangle \big)^2 \Big\rangle 
= \frac{\sqrt{2}\pi^{3/2}}{32} \frac{1}{mR} + \ldots \,.
\end{equation}

\newpage
\section{Integrals for Mass Renormalization}
\label{integrals}

For $i=1,2$,
\begin{eqnarray}
I_{i}^{(a)} & = & 
\iiint_0^1 dx\,dy\,dz\, 
\frac{\delta(x+y+z-1)}{\sqrt{xy+xz+yz}}
\int_{-\pi}^{\pi} dk
\int_{-\pi}^{\pi} dq \, \left( \frac{N_i^{(a1)}}{2(xy+xz+yz)^2 D}
+ \frac{N_i^{(a2)}}{D^2} \right) \,, \,\,
\\[5pt]
I_{i}^{(b)} & = & r 
\iiint_0^1 dx\,dy\,dz\, 
\frac{\delta(x+y+z-1)}{\sqrt{xy+xz+yz}}
\int_{-\pi}^{\pi} dk
\int_{-\pi}^{\pi} dq \, \left( \frac{N_i^{(b1)}}{2(xy+xz+yz)^2 D}
+ \frac{N_i^{(b2)}}{D^2} \right) \,, \qquad
\end{eqnarray}
where
\begin{eqnarray}
D & = & x\Big[\sin^2(k) + \Big(ma+2r\sin^2\Big(\frac{k}{2}\Big)\Big)^2\Big]
     + y\Big[\sin^2(q) + \Big(ma+2r\sin^2\Big(\frac{q}{2}\Big)\Big)^2\Big] \\
& &
     + z\Big[\sin^2(q-k) + \Big(ma+2r\sin^2\Big(\frac{q-k}{2}\Big)\Big)^2\Big] 
       - \frac{xyz}{xy+xz+yz} (ma)^2\,, \nonumber \\
N_1^{(a1)} & = & x^2 y - x y^2 + x^2z + 4 xyz - y^2 z + xz^2 + yz^2 \,, \\
N_1^{(a2)} & = & 
4f_3 r^2\sin^2\Big(\frac{k}{2}\Big)\sin^2\Big(\frac{k-q}{2}\Big)
+ f_3 \sin(k)\sin(k-q) 
+ 4f_2 r^2\sin^2\Big(\frac{k}{2}\Big)\sin^2\Big(\frac{q}{2}\Big)
\\
& & 
+  4f_1 r^2\sin^2\Big(\frac{k-q}{2}\Big)\sin^2\Big(\frac{q}{2}\Big)
- f_2\sin(k)\sin(q) + f_1\sin(k-q)\sin(q)
\nonumber \\
& & 
+ 2(ma)r \Big(
f_2f_3 \sin^2\Big(\frac{k}{2}\Big) + f_1f_3 \sin^2\Big(\frac{k-q}{2}\Big)
        + f_1f_2 \sin^2\Big(\frac{q}{2}\Big)  \Big)
+ (ma)^2 f_1f_2f_3 \,, \nonumber \\
f_1 & = & \frac{xy+xz+2yz}{xy+xz+yz} \,,\qquad
f_2 \,\, = \,\, \frac{2xy+xz+yz}{xy+xz+yz} \,,\qquad
f_3 \,\, = \,\, \frac{xy+xz}{xy+xz+yz} \,,
\nonumber \\
N_1^{(b1)} & = & x^2y-xy^2+x^2z+xyz-y^2z+xz^2+yz^2
\\
& & 
-(xy+xz+yz)(x\cos(k)-y\cos(q)+z\cos(k-q)) \,,
\nonumber \\
N_1^{(b2)} & = &  8r^2\sin^2\Big(\frac{k}{2}\Big)\sin^2\Big(\frac{k-q}{2}\Big)
\sin^2\Big(\frac{q}{2}\Big) + 2\sin(k)\sin(k-q)\sin^2\Big(\frac{q}{2}\Big)
\\
& &
-2\sin(k)\sin^2\Big(\frac{k-q}{2}\Big)\sin(q) 
+ 2\sin^2\Big(\frac{k}{2}\Big)\sin(k-q)\sin(q) \,, \nonumber \\
N_2^{(a1)} & = & 4z(xy+xz+yz) \,, \\
N_2^{(a2)} & = & 
-(ma)^2\frac{(xyz)^2}{(xy+xz+yz)^3} 
- (ma)\frac{xyz^2\Big(ma+2r\sin^2\Big(\frac{k-q}{2}\Big)\Big)}{(xy+xz+yz)^2}\\
& & 
+ \frac{2xy+xz+yz}{xy+xz+yz}\Big(\Big(ma+2r\sin^2\Big(\frac{k}{2}\Big)\Big)
\Big(ma+2r\sin^2\Big(\frac{q}{2}\Big)\Big) -\sin(k)\sin(q) \Big) 
\nonumber \\
& & 
+ 2r \sin^2\Big(\frac{k-q}{2}\Big) \Big( ma + 2r \Big( 
\sin^2\Big(\frac{k}{2}\Big) + \sin^2\Big(\frac{q}{2}\Big) \Big)\Big) \,,
\nonumber\\
N_2^{(b1)} & = & 2z(xy+xz+yz)\sin^2\Big(\frac{k-q}{2}\Big) \,, \\
N_2^{(b2)} & = & 2 \sin^2\Big(\frac{k-q}{2}\Big)
\Big(4r^2\sin^2\Big(\frac{k}{2}\Big)\sin^2\Big(\frac{q}{2}\Big)
-\sin(k)\sin(q) \Big) \,.
\end{eqnarray}

\newpage
\bibliography{fermion}

\end{document}